\documentclass[twocolumn,aps,superscriptaddress,showpacs,preprintnumbers,amsmath,amssymb]{revtex4}
\usepackage{graphicx}
\usepackage{dcolumn}
\usepackage{colordvi} 
\usepackage{bm}
\begin{document}
\title{Cosmological test of gravity with polarizations of 
  stochastic gravitational waves around 0.1-1 Hz}
\author{Atsushi Nishizawa}
\email{atsushi.nishizawa@nao.ac.jp}
\affiliation{Division of Theoretical Astronomy, National Astronomical Observatory of Japan, Mitaka, 
Tokyo 181-8588, Japan}
\author{Atsushi~Taruya}
\affiliation{Research Center for the Early Universe, School of Science, The University of Tokyo, Tokyo 113-0033, Japan}
\affiliation{Institute for the Physics and Mathematics of the Universe, University of Tokyo, Kashiwa, Chiba 277-8568, Japan}
\author{Seiji Kawamura}
\affiliation{TAMA Project, National Astronomical Observatory of Japan, Mitaka, 
Tokyo 181-8588, Japan}
\date{\today}

\begin{abstract}

In general relativity, 
a gravitational wave has two polarization modes (tensor mode), but it could have additional polarizations 
(scalar and vector modes) in the early stage of the universe, where 
the general relativity may not strictly hold and/or the effect of 
higher-dimensional gravity may become significant. 
In this paper, we discuss how to detect extra-polarization modes of 
stochastic gravitational wave background (GWB), and study the 
separability of each polarization using future space-based detectors 
such as BBO and DECIGO. We specifically consider two plausible setups of 
the spacecraft constellations consisting of two and four clusters, 
and estimate the sensitivity to each polarization mode of GWBs. 
We find that a separate detection of each polarization mode is rather sensitive to the geometric configuration and distance between clusters and that the clusters should be, in general, separated by an appropriate distance. This seriously degrades the signal sensitivity, however, 
for suitable conditions, space-based detector can separately 
detect scalar, vector and tensor modes of GWBs with energy density as
low as $h_0^2\Omega_{\rm gw}\sim10^{-15}$. 
\end{abstract}

\pacs{04.50.Kd, 04.80.Cc, 04.80.Nn.}
\maketitle

\section{Introduction}
\label{sec:intro}

Incoherent superposition of gravitational waves produced by many 
unresolved sources or diffuse sources forms a stochastic 
background of gravitational 
waves (GWs), whose statistical properties contain valuable 
information about the high-energy astrophysical phenomena 
and the cosmic structure formation. In particular, with the 
gravitational wave backgrounds (GWBs), 
we can directly probe the very early Universe 
beyond the last-scattering surface of the cosmic microwave background. 

Various mechanisms or scenarios have been proposed for 
generation of cosmological GWBs in the early universe, via the inflation \cite{bib41,bib42,bib43,bib44}, cosmological phase transition \cite{bib45,bib46,bib47,bib48}, and reheating of the Universe \cite{bib49,bib50,bib51,bib52,bib53} and etc.. An important aspect of those scenarios is that general relativity (GR) may not strictly hold in the high-energy regime of the universe, and the gravitational waves do not necessarily satisfy 
the transverse and traceless conditions. 
This implies that the number of 
polarization modes of a GW is more than that of tensor modes (i.e., 
two polarization modes called plus and cross modes), and 
it can have six modes at most in the four-dimensional spacetime, 
including scalar and vector modes \cite{bib6,bib5}. In modified gravity theories such as Brans-Dicke theory \cite{bib36,bib37} and $f(R)$ gravity \cite{bib38,bib39}, such additional polarizations appear (For more rigorous treatment of the polarizations with the Newman-Penrose formalism, see \cite{bib72,bib73}). Further, there are several attractive
scenarios that we live in a three-dimensional brane embedded in a 
higher-dimensional spacetime, such as the Kaluza-Klein theory and the Dvali-Gabadadze-Porrati (DGP) braneworld model \cite{bib40}. 
In those models, even the tensor modes 
satisfying the transverse and traceless conditions can 
have extra polarization degrees, which propagate in the extra-dimensional 
bulk spacetime. The effects of higher-dimensional gravity are expected to 
be significant at high-energy scales, and thus the
cosmological GWBs generated during such a stage may have additional 
polarization modes, which can be viewed as the mixture of 
scalar and vector polarizations in the projected three-dimensional space. 
In these respects, the polarization modes of GWBs provide 
additional information about the physics of the early universe,  
and thus a search for extra-polarization modes is indispensable as 
a cosmological test of GR. Note also that the polarization of 
GWB from astrophysical origin can also be useful as 
a test of strong gravity associated with astrophysical phenomena.

Currently, there is no observational evidence for GWBs, and the 
constraints on the extra-polarization modes of GWBs are almost nonexistent
\footnote{For constraints on the polarization of periodic GWs emitted 
from astronomical objects, the observed orbital decay of 
binary pulsars PSR B1913+16 agrees well with the prediction of GR, 
at a level of $1\,\%$ error \cite{bib13}, indicating 
that the contribution of scalar or vector GWs to the energy loss is less 
than $1\,\%$. }. 
However, the observations of 
cosmic microwave background anisotropies are currently consistent with the 
adiabatic density perturbations plus negligible contribution of tensor GWB \cite{bib61} and no significant contributions of scalar and vector GWBs 
are expected. Further, a search for stochastic GWBs by 
LIGO \cite{bib4} has given an upper limit on the energy density of 
GWBs around $\sim100$Hz, $h^2 \Omega_{\rm{gw}} \lesssim 3.6\times 10^{-6}$, 
where $\Omega_{\rm{gw}}$ is the energy density per logarithmic frequency bin 
normalized by the critical density of the Universe, and the 
present Hubble parameter normalized by 
$H_0=100\,h_0\,\rm{km\, sec^{-1}\,Mpc^{-1} }$. This null detection is 
applied to the constraints on the GWBs of extra-polarization modes 
with correction by a factor of a few, depending on the response of the GW 
detectors to each polarization mode.

In this paper, we investigate how well we can separately detect and measure 
the polarizations of a GWB using space-based GW detectors. Previously, 
we have studied the detection and separation of polarization modes of GWB
using a network of ground-based laser interferometers 
(for a detection of GWB using the pulsar timing arrays, 
see Ref. ~\cite{bib11}). With the correlation 
signals obtained from more than three advanced detectors, we found that 
scalar, vector and tensor modes of GWBs can be separately detected 
around the frequencies $f\sim100$Hz, and the sensitivity to each polarization 
mode can reach $h_0^2\Omega_{\rm{gw}}=10^{-9}\sim10^{-8}$. 
Extending the previous  
analysis to those using space-based interferometers, we discuss a
direct detection of extra-polarization modes of low-frequency GWBs at 
$f=0.1\sim1$Hz. 

\begin{figure}[t]
\begin{center}
\includegraphics[width=5cm]{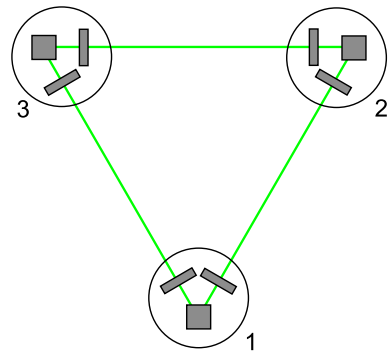}
\caption{A cluster of DECIGO.}
\label{fig10}
\end{center}
\end{figure}

Currently, several space missions to detect GWs have been 
proposed. Among these, DECI-hertz interferometer Gravitational-wave 
Observatory (DECIGO) \cite{bib7,bib10} and Big-Bang Observer (BBO) \cite{bib9} (also see \cite{bib21} for updated information) will aim at detecting 
cosmological GWBs generated during the inflationary epoch as the primary target. 
These orbit the Sun with a period of one sidereal year, and 
constitute several clusters, each of which consists of three spacecrafts  
exchanging laser beams with the others, as shown in Fig.~\ref{fig10}. 
DECIGO plans to have the arm-length $10^3\,{\rm{km}}$, 
equipped with Fabry-Perot cavity in each arm, while BBO will adopt 
the transponder type with arm-length $10^4\,{\rm{km}}$. The crucial 
difference between space- and ground-based detectors is that 
practical design as well as precise orbital configurations for 
space interferometers are still under debate, and 
there are a number of options for the detector configuration. 
Hence, in this paper, we will examine several plausible setups and 
discuss under what conditions we can separately measure the scalar, 
vector and tensor polarizations of GWB. 

This paper is organized as follows. In Sec.~\ref{sec:formulation}, 
for notational convenience, 
we first present the definitions of GW polarizations. Then, we
discuss a methodology to separately detect the polarization modes, 
based on the previous analysis using the ground-based interferometers.  
In Sec.~\ref{sec:sensitivity}, we investigate the separability of 
the polarization modes of the GWB in specific configurations 
of space-based detectors, and calculate the detector sensitivities to 
each polarization mode, especially focusing on DECIGO. 
Sec.~\ref{sec:discussion} presents
discussion on the low-frequency cutoff 
due to the presence of astronomical confusion noise, and  
the sensitivity to polarizations in the BBO case. Finally, the paper is 
summarized in Sec.~\ref{sec:summary}. 

\section{Formulation}
\label{sec:formulation}

\subsection{GW polarizations and detector response}

\begin{figure}[t]
\begin{center}
\includegraphics[width=5.5cm]{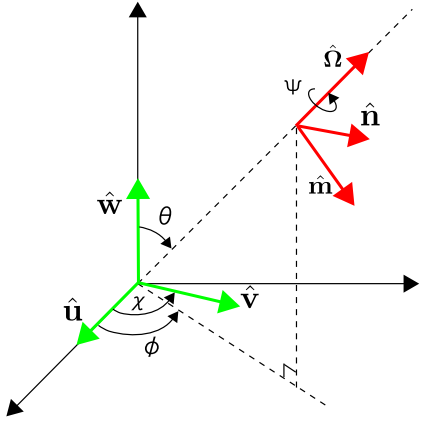}
\caption{Coordinate systems.}
\label{fig0}
\end{center}
\end{figure}

We start by briefly reviewing the basic concepts of data analysis of 
stochastic GWB search. First consider the spacetime metric generated by 
a stochastic GWB in the observed three-dimensional space.  
At a position $\vec{\mathbf{X}}$ and time $t$, it is expressed as
\begin{eqnarray}
\mathbf{h}(t,\vec{\mathbf{X}}) &=& \sum_p \int _{S^2} d\hat{\mathbf{\Omega}} \int_{-\infty}^
{\infty}df\, \nonumber \\
&& \times \tilde{h}_p (f, \hat{\mathbf{\Omega}})\, e^{2\pi if (t-\hat{\mathbf{\Omega}} \cdot 
\vec{\mathbf{X}}/c)}\, \mathbf{e}_p(\hat{\mathbf{\Omega}})\:, 
\label{eq4}
\end{eqnarray}
where $c$ is the speed of light \cite{bib19}, $f$ is frequency of a GW, and
$\hat{\mathbf{\Omega}}$ is a unit vector pointed at the GW propagating 
direction. The amplitude $\tilde{h}_p (f, \hat{\mathbf{\Omega}})$ 
represents the Fourier transform of the GW amplitude for each 
polarization mode, and the quantity $\mathbf{e}_p$ is the polarization tensor. 
Including the extra-polarization degrees 
of scalar and vector modes,  we have six polarization 
modes in three-dimensional space; $p=+, \times, b, \ell, x,$ 
and $y$, which are called plus, cross, breathing, longitudinal, vector-x, 
and vector-y modes, respectively. 
Using the orthonormal vectors, $\hat{\mathbf{m}}$ and 
$\hat{\mathbf{n}}$, perpendicular to the direction vector 
$\hat{\mathbf{\Omega}}$ (as shown in Fig. \ref{fig0}), 
the polarization tensors are defined by \cite{bib5,bib6} 
\begin{eqnarray}
\mathbf{e}_{+} &=& \hat{\mathbf{m}} \otimes \hat{\mathbf{m}} -\hat{\mathbf{n}} \otimes \hat{\mathbf{n}}\;, \nonumber \\ 
\mathbf{e}_{\times} &=& \hat{\mathbf{m}} \otimes \hat{\mathbf{n}} +\hat{\mathbf{n}} \otimes \hat{\mathbf{m}}\;, \nonumber \\
\mathbf{e}_{b} &=& \hat{\mathbf{m}} \otimes \hat{\mathbf{m}} + \hat{\mathbf{n}} \otimes \hat{\mathbf{n}}\;, \nonumber \\
\mathbf{e}_{\ell} &=& \sqrt{2}\, \hat{\mathbf{\Omega}} \otimes \hat{\mathbf{\Omega}}\;, \nonumber \\
\mathbf{e}_{x} &=& \hat{\mathbf{m}} \otimes \hat{\mathbf{\Omega}} +\hat{\mathbf{\Omega}} \otimes \hat{\mathbf{m}}\;, \nonumber \\
\mathbf{e}_{y} &=& \hat{\mathbf{n}} \otimes \hat{\mathbf{\Omega}} +\hat{\mathbf{\Omega}} \otimes \hat{\mathbf{n}}\;. \nonumber
\end{eqnarray}
Each polarization mode is orthogonal to one another and is normalized 
so that $e_{ij}^{p} e^{ij}_{p^{\prime}}=2 \delta _{pp^{\prime}}$ for 
$p,p^{\prime}=+,\times,b,\ell,x,$ and $y$. 
Note that the breathing and longitudinal modes do not satisfy the traceless condition, in contrast to the ordinary plus and cross 
polarization modes in GR. 
For the universe with extra-dimensions, the number of polarization modes generally can be more than six, but in 
the projected three-dimensional space, GW can be viewed as a mixture 
of scalar, vector and tensor modes mentioned above.

Next consider the response of the GW detector. 
The laser interferometers measure the time variation of 
the spacetime metric as one-dimensional time-series data. In a space-based interferometer, the gravitational-wave signal is obtained by differentiating two link signals in Fig.\ref{fig10} (three interferometer signals is obtained about a cluster.). Denoting the signal strain measured by the interferometer $I$ 
(whose position is located at $\vec{\mathbf{X}}$) by $h_I(t)$, 
the strain amplitude of GW is expressed as
\begin{eqnarray}
h_I(t,\vec{\mathbf{X}}) &=& \mathbf{D}_I : \mathbf{h}(t,\vec{\mathbf{X}}) \nonumber \\
&=& \sum_p \int _{S^2} d\hat{\mathbf{\Omega}} \int_{-\infty}^{\infty}df\, \nonumber \\
&& \times \,\tilde{h}_p (f, \hat{\mathbf{\Omega}})\,e^{2\pi if (t-\hat{\mathbf{\Omega}} \cdot \vec
{\mathbf{X}}/c)} F_I^p(f, \hat{\mathbf{\Omega}})\:, 
\label{eq7}
\end{eqnarray}
where the quantity $\mathbf{D}_I$ is the detector tensor, and the 
$F_I^p$ is the angular response function for each polarization mode. They 
are respectively given by 
\begin{eqnarray}
F^p_I (\hat{\mathbf{\Omega}}) &\equiv & \mathbf{D}_I : \mathbf{e}_p
(\hat{\mathbf{\Omega}})\;, 
\label{eq5} \\
\mathbf{D}_I &\equiv & \frac{1}{2}\left[ \hat{\mathbf{u}} \otimes \hat{\mathbf{u}} - \hat{\mathbf{v}}
\otimes \hat{\mathbf{v}} \right]\:, 
\label{eq1}
\end{eqnarray}
with the unit vectors $\hat{\mathbf{u}}$ and $\hat{\mathbf{v}}$ being 
directed to each detector arm. The expression of Eq. (\ref{eq1}) is valid 
when the arm length of the detector, $L$, is much smaller than the 
wavelength of observed GWs, $\lambda_{\rm{g}}$, i.e., $L\ll\lambda_{\rm g}$. 
For DECIGO, the observable frequency range is around $f\sim 1\,\rm{Hz}$, 
which corresponds to $\lambda_{\rm{g}}=3\times 10^5\,\rm{km}$. 
Thus, with the arm length $L=10^3\,\rm{km}$, the  
so-called {\it{low frequency approximation}} is fully satisfied.

\subsection{Cross-correlation analysis}

Throughout the paper,  
we assume that stochastic GWB is (i) isotropic, 
(ii) stationary, (iii) Gaussian, and (iv) has no intrinsic correlation between polarization modes (If this is not the case, see \cite{bib62,bib63,bib64,bib65,bib66,bib67,bib68} and \cite{bib69,bib70,bib71} for discussions on the detection of GWBs in the presence of anisotropies and non-Gaussianity, respectively.). Adopting these assumptions, 
all the statistical properties of the GWB are characterized by the 
power spectral density: 
\begin{eqnarray}
\langle \tilde{h}_p^{\ast} (f,\hat{\Omega} ) \tilde{h}_{p^{\prime}} (f^{\prime},\hat{\Omega^
{\prime}} ) \rangle &=& \delta(f-f^{\prime})\frac{1}{4\pi} \delta^2 ( \hat{\Omega},\hat
{\Omega^{\prime}}) \delta_{pp^\prime} \nonumber \\
&& \quad \quad \times \, \frac{1}{2} S_h^p (|f|)\;,
\label{eq21}
\end{eqnarray}
where $\delta^2 ( \hat{\Omega},\hat{\Omega^{\prime}}) 
\equiv \delta(\phi-\phi^{\prime}) \delta (\cos\theta - \cos\theta^\prime )$, 
and $\langle \cdots \rangle$ denotes ensemble average. The function 
$S_h^p(f)$ is the one-sided power spectral density for each polarization mode.

Conventionally, the amplitude of GWB for each polarization is also 
characterized by an energy density per logarithmic frequency bin, 
normalized by the critical energy density of the Universe:
\begin{equation}
\Omega_{\rm{gw}}^p (f) \equiv \frac{1}{\rho_c}\frac{d\rho_{\rm{gw}}^p}{d \ln f} 
= \left( \frac{2 \pi^2}{3H_0^2} \right) f^3 S_h^p (f) \:, 
\label{eq22}
\end{equation}
where $\rho_c = 3 H_0^2/8\pi G$ and $H_0$ is the Hubble constant. In the 
second equality, we used the relation between $\Omega_{\rm{gw}}(f)$ and 
$S_h(f)$ given by \cite{bib3,bib27}. Then, we define the GWB energy density 
in tensor, vector, and scalar polarization modes as
\begin{eqnarray}
\Omega_{\rm{gw}}^T &\equiv & \Omega_{\rm{gw}}^{+} + \Omega_{\rm{gw}}^{\times} \;, 
\nonumber \\
\Omega_{\rm{gw}}^V & \equiv& \Omega_{\rm{gw}}^x + \Omega_{\rm{gw}}^y \;, 
\nonumber \\
\Omega_{\rm{gw}}^S &\equiv & \Omega_{\rm{gw}}^b + \Omega_{\rm{gw}}^{\ell} = 
\Omega_{\rm{gw}}^b (1+\kappa) \;, \nonumber 
\end{eqnarray}
The subscripts $T$, $V$, and $S$ stand for tensor, vector, and scalar, 
respectively. Hereafter, 
we assume $\Omega_{\rm{gw}}^{+} = \Omega_{\rm{gw}}^{\times}$ 
for the tensor mode and $\Omega_{\rm{gw}}^{x} = \Omega_{\rm{gw}}^{y}$ for the 
vector mode. This assumption is valid for a stochastic GWB generated in 
most of cosmological scenarios \cite{bib1}. For the scalar mode, we 
introduce a model-dependent new parameter, 
$\kappa(f) \equiv \Omega_{\rm{gw}}^{\ell}(f)/\Omega_{\rm{gw}}^{b}(f)$.

In order to discriminate a stochastic GWB from random detector noise, 
one needs to cross-correlate between detector signals 
\cite{bib25,bib26,bib27}. Let us 
consider the outputs of a detector, $s(t)=h(t)+n(t)$, where $h(t)$ and 
$n(t)$ are the GW signal and the noise of a detector. In general, the 
amplitude of GWB is thought to be much smaller than that of detector noise. 
Cross-correlation signal $Y$ between two detectors is given by
\begin{eqnarray}
Y &\equiv& \int_{-T_{\rm{obs}}/2}^{T_{\rm{obs}}/2} dt \int_{-T_{\rm{obs}}/2}^{T_{\rm{obs}}/2} dt^{\prime}\, s_I(t) s_J(t^{\prime} ) Q(t-t^
{\prime})\;, \nonumber \\
&\approx&\int_{-\infty}^{\infty}df \int_{-\infty}^{\infty}df^{\prime} \delta _T (f-f ^{\prime}) \tilde{s}_I^
{*}(f) \tilde{s}_J(f^{\prime}) \tilde{Q}(f^{\prime}), 
\label{eq20}
\end{eqnarray}
where $T_{\rm{obs}}$ is observation time, $\tilde{s}_I(f)$, $\tilde{s}_J(f)$ 
and $\tilde{Q}(f)$ are the Fourier transforms of 
$s_I(t)$, $s_J(t)$ and $Q(t-t^{\prime})$, respectively. 
$Q(t-t^{\prime})$ is a filter function, which will be later adjusted to 
maximize the signal-to-noise ratio (SNR) of correlation signal.  
The function $\delta_T(f)$ is defined by
\begin{equation}
\delta _T (f)\equiv \int_{-T_{\rm{obs}}/2}^{T_{\rm{obs}}/2} dt \,e^{-2\pi ift}=\frac{\sin(\pi fT_{\rm{obs}})}{\pi f}\:.
\nonumber
\end{equation}

Taking the ensemble average over the expression (\ref{eq20}), 
and substituting Eqs. (\ref{eq21}) and (\ref{eq22}) into this, 
we obtain a GW signal in a correlation analysis between $I$-th and $J$-th 
detectors, 
\begin{eqnarray}
\mu &\equiv& \langle Y \rangle \nonumber \\
&=&\int_{-\infty}^{\infty}df \int_{-\infty}^{\infty}df^{\prime} \delta_T(f-f^{\prime}) \langle \tilde{h}_I^{*}(f) \tilde{h}_J(f^{\prime}) 
\rangle \tilde{Q}(f^{\prime})\:. \nonumber \\
&=& \frac{3H_0^2}{20\pi^2} T_{\rm{obs}}\,\sin ^2 \chi \int_{-\infty}^{\infty} df |f|^{-3} \tilde{Q}(f) \nonumber \\
& \times & \biggl[ \Omega_{\rm{gw}}^T(f) \gamma^T_{IJ}(f) + \Omega_{\rm{gw}}^V(f) \gamma^V_{IJ}(f) \nonumber \\
&& +\xi(f) \, \Omega_{\rm{gw}}^S(f) \gamma^S_{IJ}(f) \biggr]  \;, 
\label{eq23} \\
\xi (f) &\equiv & \frac{1}{3} \left( \frac{1+2\kappa (f)}{1+\kappa (f)} \right) \;. \nonumber 
\end{eqnarray}
The parameter $\xi$ takes the value in the range $1/3 \leq \xi \leq 2/3$ 
and characterizes the ratio of the energy in the longitudinal mode to the 
breathing mode. The sensitivity to the GWB can be 
governed by the so-called overlap reduction functions \cite{bib25,bib26,bib27}, which represents how much of the correlation of the GW signal 
between detectors can be preserved. The overlap reduction function for each polarization is defined by \cite{bib12} 
\begin{eqnarray}
\gamma_{IJ}^{M}(f) &\equiv& \frac{1}{\sin^2 \chi} \int_{S^2} \frac{d\hat{\Omega}}{4 \pi}\, e^{2\pi i f \hat{\Omega}\cdot \Delta \vec{X}/c}\, {\cal{R}}_{IJ}^M \;,
\label{eq6} \\
{\cal{R}}_{IJ}^T(\hat{\mathbf{\Omega}}) &\equiv & \frac{5}{2}\, (F_I^{+} F_J^{+} + F_I^{\times} F_J^{\times})\;, \nonumber \\
{\cal{R}}_{IJ}^V(\hat{\mathbf{\Omega}}) &\equiv & \frac{5}{2}\, (F_I^{x} F_J^{x} + F_I^{y} F_J^{y})\;, \nonumber \\
{\cal{R}}_{IJ}^S(\hat{\mathbf{\Omega}}) &\equiv & \frac{15}{1+2\kappa}\, (F_I^{b} F_J^{b} +\kappa F_I^{\ell} F_J^{\ell})\;, \nonumber 
\end{eqnarray}
which are normalized to unity in the limit $f\to0$. 
The subscript $M$ denotes $M=T,V,S$, and the quantity
$\Delta \vec{\mathbf{X}}$ is the separation vector defined by 
$\Delta \vec{\mathbf{X}}\equiv \vec{\mathbf{X}}_I-\vec{\mathbf{X}}_J$. 
Note that the prefactor, 
$\sin^2 \chi = 1-(\hat{\mathbf{u}} \cdot \hat{\mathbf{v}})^2$,   
in Eq.~(\ref{eq6}) comes from the non-orthogonal detector arms. 
For an equilateral triangle configuration of the spacecraft constellation 
in Fig.~\ref{fig10}, we have $\sin^2 \chi = 3/4$.

In Eq.~(\ref{eq6}),  the angular integral 
is analytically performed prior to specifying 
the detector location and orientation \cite{bib12}. The result is expressed as
\begin{eqnarray}
\gamma_{IJ}^{M} (f) &=& \frac{1}{\sin^2 \chi} \biggl[ \rho_1^{M} (\alpha) D_I^{ij} D^J_{ij} + \rho_2^{M} (\alpha) D_{I,\,k}^{i} D_J^{kj} \hat{d}_i \hat{d}_j \nonumber \\
&&+\rho_3^{M} (\alpha) D_I^{ij} D_J^{k\ell} \hat{d}_i \hat{d}_j \hat{d}_k \hat{d}_{\ell}
 \biggr] \;, 
 \label{eq25}
\end{eqnarray}  
with unit vector $\hat{d}_i$ defined by 
$\hat{d}_i \equiv \Delta \vec{X}/ |\Delta \vec{X}|$. 
The summation is taken over each component of the subscripts $i,j,k,\ell$. 
In the above, 
frequency dependence of the overlap reduction function is 
incorporated into the coefficients, $\rho^M_1$, $\rho^M_2$, and $\rho^M_3$, 
which sensitively depend on each polarization mode. 
We have
\begin{equation}
\left(
\begin{array}{c} 
\rho_1^{T} \\
\rho_2^{T} \\
\rho_3^{T} 
\end{array}
\right)
=\frac{1}{14}
\left(
\begin{array}{ccc} 
28 & -40 & 2  \\
0 & 120 & -20 \\
0 & 0 & 35
\end{array}
\right)
\left(
\begin{array}{c} 
j_0 \\
j_2 \\
j_4  
\end{array}
\right)\;, \nonumber
\end{equation}
for tensor mode, 
\begin{equation}
\left(
\begin{array}{c} 
\rho_1^{V} \\
\rho_2^{V} \\
\rho_3^{V} 
\end{array}
\right)
=\frac{2}{7}
\left(
\begin{array}{ccc} 
7 & 5 & -2  \\
0 & -15 & 20 \\
0 & 0 & -35
\end{array}
\right)
\left(
\begin{array}{c} 
j_0 \\
j_2 \\
j_4  
\end{array}
\right)\;, \nonumber
\end{equation}
for vector mode, and 
\begin{equation}
\left(
\begin{array}{c} 
\rho_1^{S} \\
\rho_2^{S} \\
\rho_3^{S} 
\end{array}
\right)
=\frac{1}{7}
\left(
\begin{array}{ccc} 
14 & 20 & 6 \\
0 & -60 & -60 \\
0 & 0 & 105
\end{array}
\right)
\left(
\begin{array}{c} 
j_0 \\
j_2 \\
j_4  
\end{array}
\right)\;. \nonumber
\end{equation} 
for scalar mode. 
Here, $j_n (\alpha)$ is the spherical Bessel function 
with its argument given by
\begin{equation}
\alpha(f) \equiv \frac{2 \pi f D}{c}\;, \quad \quad D \equiv |\Delta \vec{X}|\;. 
\label{eq26}
\end{equation}
These expressions are very useful to obtain a simple expression 
for the overlap reduction function in specific detector 
configurations below.

\begin{figure}[t]
\begin{center}
\includegraphics[width=8cm]{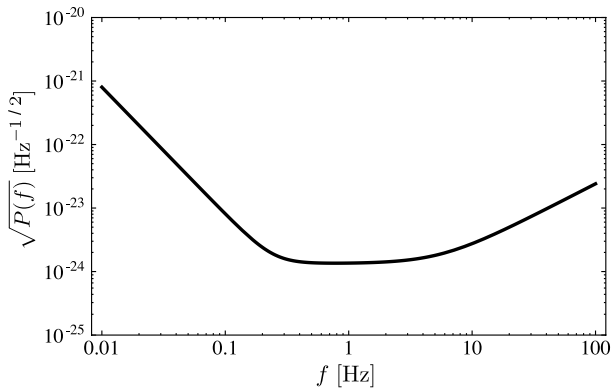}
\caption{DECIGO noise curve.}
\label{fig5}
\end{center}
\end{figure}

Let us now consider the noise part in the cross correlation analysis. 
As long as the intrinsic noise correlation between two detectors is 
absent, the ensemble average of cross correlation quantity $Y$ in 
Eq.~(\ref{eq20}) is dominated by the GW signals. This is true even 
in the weak signal limit, $h\ll n$. 
However, the variance of $Y$ is dominated by the detector noises. We obtain
\begin{eqnarray}
\sigma ^2 &\equiv & \langle Y^2 \rangle - \langle Y \rangle ^2 \nonumber \\
&\approx & \langle Y^2 \rangle \nonumber \\
&\approx& \int_{-\infty}^{\infty} df \int_{-\infty}^{\infty} df^{\prime}\, \tilde{Q}(f) \tilde{Q}^{\ast}(f^{\prime})\, \nonumber \\
&& \times \langle 
\tilde{n}^{\ast}_I(f) \tilde{n}_I(f^{\prime}) \rangle \, \langle \tilde{n}_J(f) \tilde{n}^{\ast}_J
(f^{\prime}) \rangle \nonumber \\
&\approx &\frac{T_{\rm{obs}}}{4} \int_{-\infty}^{\infty} df \,P_I(|f|) P_J(|f|)\, | \tilde{Q}(f) | ^2\:, 
\label{eq24}
\end{eqnarray}
where the one-sided power spectrum density of the detector noise, $P_I(f)$, 
is defined by
 \begin{equation}
\langle \tilde{n}^{\ast}_I(f) \tilde{n}_J(f^{\prime}) \rangle \equiv \frac{1}{2}\delta (f-f^
{\prime}) \delta_{IJ} P_I(|f|)\;. \nonumber
\end{equation}
For DECIGO, the analytical fit of the noise power spectrum is obtained 
for a single interferometer. Assuming that the detector noise 
is idealistically limited by the sum of quantum noises, i.e., 
shot noise, $P^{\rm{shot}}$, 
and radiation-pressure (acceleration) noise, $P^{\rm{acc}}$, we have 
\cite{bib16}: 
\begin{eqnarray}
P (f)&=&P^{\rm{acc}} (f)+P^{\rm{shot}} (f) \nonumber \\
P^{\rm{acc}} (f) &=& 6.31\times 10^{-51} \biggl(\frac{f}{1\rm{Hz}} \biggr)^{-4} \; \rm{Hz}^{-1} \;, \nonumber \\
P^{\rm{shot}} (f) &=& 1.88\times 10^{-48} +5.88\times 10^{-50} \biggl(\frac{f}{1\rm{Hz}} \biggr)^{2} \; \rm{Hz}^{-1} \;. \nonumber
\end{eqnarray}
In Fig.~\ref{fig5}, the noise power spectrum of DECIGO is plotted as 
the strain amplitude, $S_h^{1/2}$.

From Eqs. (\ref{eq23}) and (\ref{eq24}), the SNR in the 
correlation analysis between two detectors is simply 
given by ${\rm{SNR}}=\mu /\sigma$. In the absence of extra-polarization
degrees (i.e., only the tensor modes exit), two-detector correlation 
is sufficient to detect GWB, and the optimal choice of 
the filter function $\tilde{Q}(f)$ is 
easy to derive \cite{bib27}. On the other hand, in the presence of 
multiple polarization modes, we need more than three detectors 
in order to separately detect each polarization mode. The optimal 
SNR combining multiple detectors is not simply given by the 
sum of $\mu /\sigma$, and  thus the choice of filter function 
is rather non-trivial. We will discuss this issue in the next section.

\subsection{Signal-to-noise ratio}
\label{sec2c}

In principle, 
three polarization modes, i.e., scalar, vector and tensor, can be separately
detected by linearly combining more than three independent correlation signals. 
In our previous paper \cite{bib12}, we considered the situation that 
only the three correlation signals are available. We then presented the formula 
for optimal SNR. Here,  we consider the optimal SNR 
combining arbitrarily large number, $N_{\rm{pair}}$, of correlation signals. 
Such a generalized formula has been derived for the cases 
with two polarizations 
(i.e., circularly polarized and un-polarized modes of tensor GWs) 
by Seto and Taruya \cite{bib14}. Based on this, in Appendix \ref{appA}, 
the extension of the formula to the three-polarization case is presented. 
Combining $N_{\rm{pair}}$ correlation signals,  
the resultant optimal SNR for separately detecting scalar, vector and tensor  
GWBs becomes (see Eq. (\ref{eq47})) 
\begin{eqnarray}
{\rm{SNR}}^M = \frac{9 H_0^2}{40 \pi^2} \biggl[2 \int_{0} ^{\infty} df \,\frac{(\Omega_{\rm{gw}}^M (f))^2 \det \mathbf{F} (f)}{f^6 {\cal{F}}_M(f)} \biggr] ^{1/2}, \;\; 
\label{eq2} \\
\nonumber \\
\mathbf{F} (f) = \left(
\begin{array}{ccc} 
F_{TT} & F_{TV} & F_{TS} \\
F_{TV} & F_{VV} & F_{VS} \\ 
F_{TS} & F_{VS} & F_{SS}
\end{array}
\right)\;, \quad \quad \quad \quad \quad \;\; \nonumber \\
\nonumber \\
F_{MM^{\prime}}(f) = \sum_i \int_{0}^{T_{\rm{obs}}} dt \, \frac{\gamma_i^M (t,f) \gamma_i^{M^{\prime}} (t,f)}{{\cal{N}}_i(f) } \;, \quad \;\; 
\label{eq3} 
\end{eqnarray}
where $M$ and $M^{\prime}$ denote polarization modes, $M,M^{\prime}=T,V,S$. 
The quantity ${\cal{F}}_M$ is the determinant of the sub-matrix, which is 
constructed by removing the $M$'s elements from $\mathbf{F}$. The 
subscript $i$ indicates a pair of detectors (e.g., $i=(I,J)$
for pair of $I$- and $J$-th detectors), 
and ${\cal{N}}_{i}(f)$ is defined as, say, 
${\cal{N}}_{12}(f) \equiv P_1(f) P_2(f)$. In what follows, for 
simplicity, we consider the case that all interferometers have 
the same noise spectrum, i.e., $P_I(f)=P(f)$.

The expression (\ref{eq2}) is a rather general formula for optimal SNR 
in the sense that the stationary configuration of GW detectors is not strictly 
assumed. The configuration of space-based interferometers gradually changes in time due to the orbital 
motion of the spacecrafts. In the formula (\ref{eq2}), 
the effect of such gradual change is incorporated into the explicit 
time dependence of the overlap reduction function, which will be important later in Sec.~\ref{subsec:case2}. 
Note that for stationary detector configuration, 
the time integral in Eq. (\ref{eq3}) simply reduces to the factor 
$T_{\rm{obs}}$, which gives rise to the well-known result that 
$SNR\propto T_{\rm{obs}}^{1/2}$. Further, if we consider
the combination of three detectors ($N_{\rm{pair}}=3$), 
the above SNR can be reduced to Eq. (53) 
in Ref.~\cite{bib12} except for the prefactor 
$\sin^2 \chi = 3/4$ or (\ref{eq8}) in this paper.

The expression (\ref{eq2}) is one of the most important results of this paper. 
Provided the location and orientation of GW detectors, 
the SNR is quantitatively evaluated. 
Before doing so, it is important to note that 
a separate detection of three polarization modes is possible only
when the quantity $\det \mathbf{F}$ becomes non-vanishing. 
As we explain in detail below, this implies that the configuration of space-based detectors 
must satisfy both conditions: 
\begin{description}
\item[(i)] The detectors have to be separated by at least the distance of a typical wavelength of the observed GWs, e.g., 
$3\times 10^5\,{\rm{km}}$ for a GW with frequency $f=1\,{\rm{Hz}}$. 
\item[(ii)] Detector pairs are not geometrically 
degenerate, e.g., three detectors located at the vertices of a 
non-equilateral triangle. 
\end{description}
If one of the two conditions fails, 
the SNR is expected to be significantly degraded.

For intuitive explanation of the above two conditions, let us consider the three-detector (three correlation-signal) case \cite{bib12}  (This does not lose generality because one can verify that SNR with an arbitrary number of detectors can be reduced to weighted sum of SNR with three-detector subset.). In the case, the condition $\det \mathbf{F} \neq 0$ corresponds to $\det \mathbf{\Pi} \neq 0$ where   
\begin{equation}
\mathbf{\Pi}(f) \equiv
\left(
\begin{array}{ccc} 
\gamma^T_{12} & \gamma^V_{12} & \gamma^S_{12} \\
\gamma^T_{23} & \gamma^V_{23} & \gamma^S_{23} \\
\gamma^T_{31} & \gamma^V_{31} & \gamma^S_{31}
\end{array}
\right) \;. \nonumber
\end{equation} 
The condition (i) comes from nondegeneracy of the components in a column of $\mathbf{\Pi}$, e.g. $\gamma^T_{12} \neq \gamma^V_{12} \neq \gamma^S_{12}$. As we will see explicitly in Sec. \ref{sec:sensitivity}, for a closer detector pair ($\alpha \rightarrow 0$), there is no difference in the overlap reduction functions for each polarization mode, since the spherical Bessel functions, $j_2$ and $j_4$ vanish. On the other hand, for a detector pair with $\alpha \sim 1$, $j_2$ and $j_4$ are of the same order as $j_0$ and result in differences between the overlap reduction functions. The condition (ii) comes from nondegeneracy of the components in a row of $\mathbf{\Pi}$, e.g. $\gamma^T_{12} \neq \gamma^T_{23} \neq \gamma^T_{31}$. This condition implies that a non-colinear configuration of three detectors is preferred.

\section{Sensitivity to polarization modes}
\label{sec:sensitivity}

We are in a position to discuss how well one can separately detect 
scalar, vector and tensor GWBs. In this section,  
we specifically consider the two setups of detector configuration, and 
estimate the detectability for each polarization mode. 
First, we consider the four-cluster configuration with coplanar orbits 
(case I). This is the prototypical configuration proposed at  
an early phase of the conceptual design of DECIGO \cite{bib10}. 
We then move to a discussion of two-cluster configuration, in which the orbits of two clusters 
are slightly inclined in relation to one another (case II). 
In the calculations below, the energy spectrum of GWBs 
$\Omega_{\rm{gw}}^M (f)$ is assumed to be a flat spectrum, i.e., 
$\Omega_{\rm{gw}}^M=$const. In computing SNR below, we 
do not consider the single-cluster correlation.  
This is because correlation signals from a single cluster are not sensitive enough to GWB at low frequencies, as discussed in Appendix \ref{appB}. 

\begin{figure}[t]
\begin{center}
\includegraphics[width=8cm]{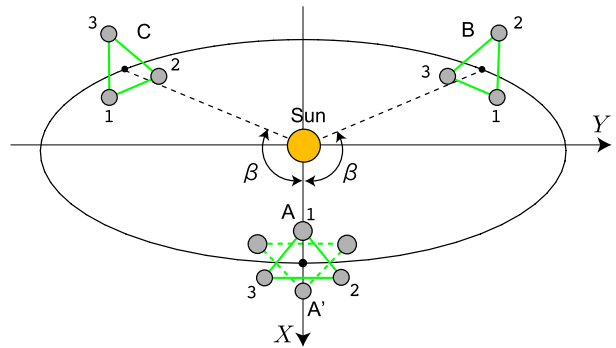}
\caption{Case I --- Four clusters, A, A', B, and C, 
sharing the coplanar orbit, whose radius is $1\,{\rm{AU}}$.}
\label{fig1}
\end{center}
\end{figure}

\subsection{Case I: four clusters}

\subsubsection{Configuration}

Let us consider the detector configuration consisting of four clusters 
shown in Fig. \ref{fig1}. 
Each cluster is inclined by 60 degrees from the orbital plane in 
order to close the orbit. 
The guiding center of each cluster, i.e., 
the center-of-mass of three spacecrafts, 
follows a circular orbit around the Sun, with the radius 
$R_0=1\,{\rm{AU}}\approx 1.5\times 10^8\,{\rm{km}}$ 
and orbital period of one year. In the coordinate system 
$(X, Y, Z)$ shown in Fig. \ref{fig1}, the position of the guiding 
center is given by
\begin{equation}
\vec{X} (t) =  (R_0\cos \phi (t),\, R_0\sin \phi (t) ,\,0\, ) \;, \nonumber
\end{equation}
where the phase of the orbit is 
$\phi (t)=\omega_{\rm{orbit}} t =2\pi (t/{1\rm{yr}})$ 
for the clusters A and A'. The phases of 
the clusters B and C are relatively shifted to $\beta$ 
and $-\beta$ from that of the clusters A and A', respectively. 
Thus, the distances between the clusters, 
$D \equiv |\Delta \vec{\mathbf{X}}|$, become 
$D_{\rm{AB}}(\beta)=D_{\rm{AC}}(\beta)=2R_0 |\sin (\beta/2)|$ for the 
AB and AC link, and $D_{\rm{BC}}(\beta)=2R_0 |\sin \beta|$ for the BC link.

In each cluster, 
the position of the spacecraft relative to the guiding center is given by
\begin{eqnarray}
\vec{\mathbf{x}}_i (t) &=& 
\mathbf{R}_Z \bigl[ -\phi(t) \bigr] \,
\mathbf{R}_Y \bigl[ -\theta \bigr]\, \mathbf{R}_Z 
\bigl[ \phi(t) \bigr]\, \vec{\mathbf{x}}_{i,0} \;, 
\label{eq37} \\
\vec{\mathbf{x}}_{i,0} &=& \frac{L}{\sqrt{3}}\times 
(\,\cos \sigma_i,\, \sin \sigma_i ,\,0 \,) \;. 
\label{eq38}
\end{eqnarray}
with $\theta=\pi/3$.  
The matrices $\mathbf{R}_Y$ and $\mathbf{R}_Z$ are the 
rotation matrices around 
$Y$ and $Z$ axes, respectively. The angle $\sigma_i$ represents 
the orientation 
angle of the bisector of two arms of each interferometer. 
Let the angle of the interferometer 1 be $\sigma_1=\sigma_0$. 
The orientations of interferometer 2 and 3 in a cluster are given 
by $\sigma_2=\sigma_0+2\pi/3$ and $\sigma_3=\sigma_0+4\pi/3$, respectively.

In the setup mentioned above, the angles $\beta$ and $\sigma_0$ are 
apparently regarded as the free parameters. However, 
in an optimal combination of detector signals, several examinations 
reveal that the resultant SNR is 
turns out to be insensitive to any choice of $\sigma_0$. We thus 
set $\sigma_0=0$, and treat the separation angle $\beta$ as the only 
free parameter. For simplicity of the calculation below, the separation 
of the detectors between different clusters is approximated as 
the distance between the guiding centers of each cluster. This treatment 
is validated as long as we are interested in the low-frequency GWs 
satisfying $\lambda_g \gg L$. Then, the detector configuration 
can become stationary, and no explicit time-dependence appears at the  
overlap reduction functions in Eq. (\ref{eq2}).

\subsubsection{Overlap reduction functions}

For each detector link of four-cluster configuration, 
the overlap reduction function 
given by Eq. (\ref{eq25}) is reduced to a rather 
compact expression. For correlation signals of AB, AC, BC links, 
the overlap reduction functions are respectively written as 
\begin{eqnarray}
\gamma^M_{AB}(f,\beta,\sigma_A,\sigma_B; \alpha_{AB}) = \frac{1}{16} \quad \quad \quad \quad \quad \quad \quad \quad \quad \nonumber \\
\times \biggl[ \Theta_1^M (\alpha_{AB}, \beta/2)\, \cos[2(\sigma_A-\sigma_B)] \quad \quad \quad \quad \nonumber \\ 
- \Theta_2^M (\alpha_{AB}, \beta/2)\, \sin [2(\sigma_A-\sigma_B)] \quad \quad \quad \quad \nonumber \\
+\Theta_3^M (\alpha_{AB}, \beta/2)\, \cos[2\beta - 2(\sigma_A+\sigma_B)] \biggr] \;, \;\;
\label{eq27} 
\end{eqnarray}
\begin{eqnarray}
 \gamma^M_{AC}(f,\beta,\sigma_A,\sigma_C; \alpha_{AC}) = \gamma^M_{AB}(f,-\beta,\sigma_A,\sigma_C; \alpha_{AB}) \;,  \nonumber \\
\end{eqnarray}
and
\begin{eqnarray} 
\gamma^M_{BC}(f,\beta,\sigma_B,\sigma_C; \alpha_{BC}) = \frac{1}{16} \quad \quad \quad \quad \quad \quad \quad \quad \quad \nonumber \\
\times \biggl[ \Theta_1^M (\alpha_{BC}, \beta)\, \cos[2(\sigma_B-\sigma_C)] \quad \quad \quad \quad \quad \nonumber \\
 + \Theta_2^M (\alpha_{BC}, \beta)\, \sin [2(\sigma_B-\sigma_C)] \quad \quad \quad \quad \quad \nonumber \\
 +\Theta_3^M (\alpha_{BC}, \beta)\, \cos[ 2(\sigma_B+\sigma_C)] \biggr] \;. \quad \quad \quad \;\;\;  
\label{eq28} 
\end{eqnarray}
Note that the overlap reduction function for 
$AA'$ link becomes $\gamma^M_{AA'}=1$, because of the mirror symmetry.  
As for A'B and A'C links, the overlap reduction functions are 
identical to those for AB and AC links. 
Here, the function $\Theta^M_i(\alpha,\beta)$ is defined by
\begin{eqnarray}
\Theta_{1,2}^M (\alpha, \beta) \equiv J_M(\alpha) \,U_{1,2}(\beta) \;, \nonumber 
\end{eqnarray}
\begin{eqnarray}
\left(
\begin{array}{c} 
\Theta_3^T (\alpha, \beta) \\ \Theta_3^V (\alpha, \beta) \\ \Theta_3^S (\alpha, \beta)  
\end{array}
\right) 
& \equiv & -9 \left[ 
\left(
\begin{array}{c} 
J_T(\alpha) \\ J_V(\alpha) \\ J_S(\alpha) 
\end{array}
\right) \sin^4 \beta \right. \nonumber \\
&+& \frac{5}{14} 
\left(
\begin{array}{c} 
-8 j_2 (\alpha)-j_4 (\alpha) \\ 4 j_2 (\alpha)+4j_4 (\alpha) \\ 8 j_2 (\alpha)-6j_4 (\alpha) 
\end{array}
\right) \sin^2 \beta \nonumber \\
&+& \left. \frac{5}{9} 
\left(
\begin{array}{c} 
j_4 (\alpha) \\ -4 j_4 (\alpha) \\ 6 j_4 (\alpha) 
\end{array}
\right)
 \right] \;. \nonumber
\end{eqnarray} 
\begin{eqnarray}
U_1(\beta) &\equiv& \sin^4 \beta +8 \cos^6 \beta (1+\cos^2 \beta ) \;, 
\label{eq29} \\
U_2(\beta) &\equiv& 2 \bigl( 1+ \cos^2 \beta \bigr) \bigl( 1+ \cos^2 \beta +2\cos^4 \beta  \bigr)\, \sin 2\beta \;,  \nonumber \\
&& \\
J_T (\alpha) &\equiv& j_0 (\alpha) +\frac{5}{7} j_2 (\alpha) + \frac{3}{112} j_4 (\alpha) \;, 
\\
J_V (\alpha) &\equiv& j_0 (\alpha) -\frac{5}{14} j_2 (\alpha) - \frac{3}{28} j_4 (\alpha) \;, \\
J_S (\alpha) &\equiv& j_0 (\alpha) -\frac{5}{7} j_2 (\alpha) + \frac{9}{56} j_4 (\alpha) \;. 
\label{eq30} 
\end{eqnarray}
The subscript $M$ stands for the polarization mode, $T, V, S$. 
The angles $\sigma_A$, $\sigma_B$, and $\sigma_C$ are the orientation of 
the interferometer in the clusters A, B, and C, respectively. $\alpha_{AB}$, $\alpha_{AC}$, and $\alpha_{BC}$ imply 
the dimensionless frequency $\alpha$ defined in Eq. (\ref{eq26}) 
for AB, AC, and BC links. 
Note that from the expressions (\ref{eq27})-(\ref{eq28}), 
the overlap reduction functions are invariant under the transformation 
$\sigma\to\sigma+n\pi/2$ except for an overall sign. This is due to the quadrupole nature of GWs.

\begin{figure*}[t]
\begin{center}
\includegraphics[width=14cm]{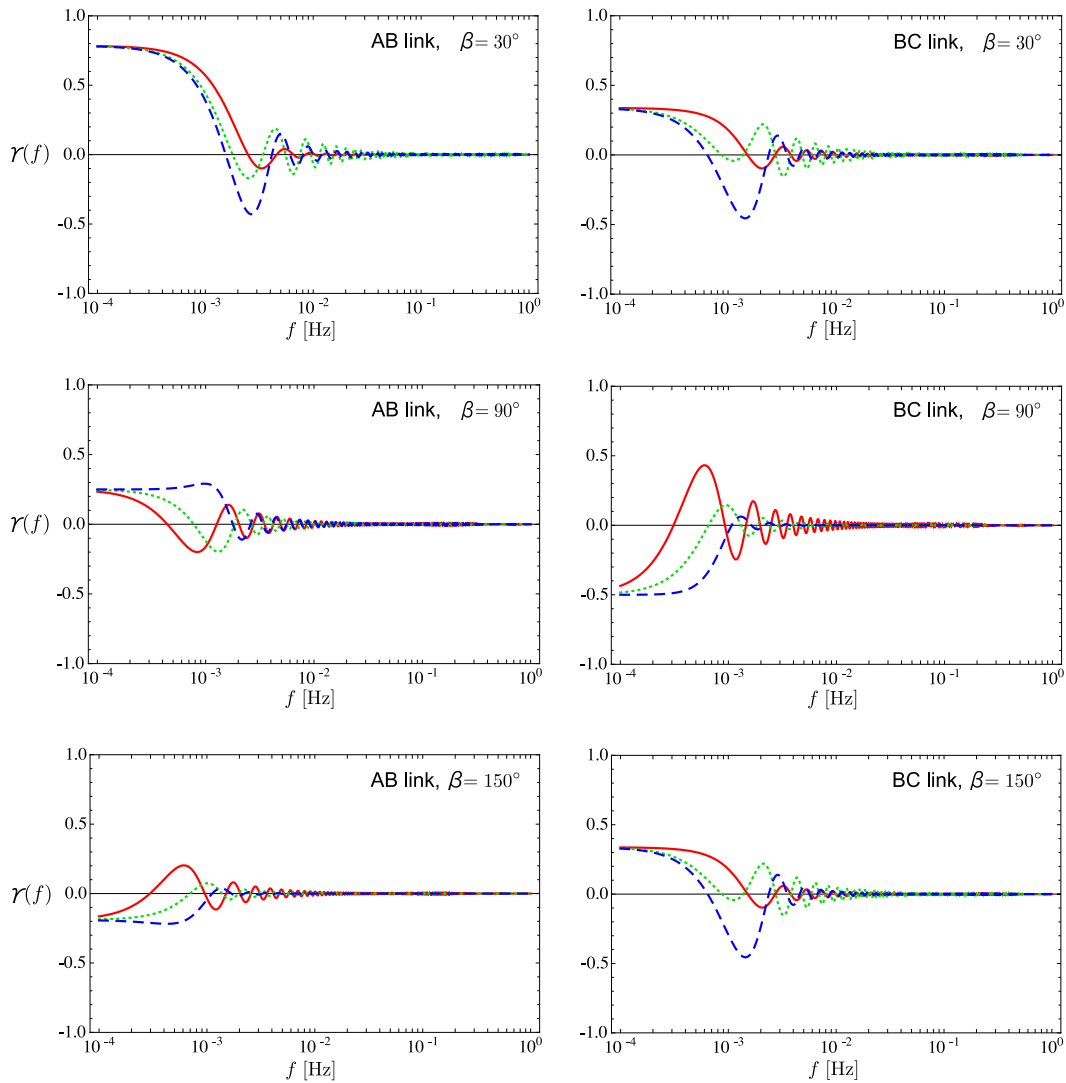}
\caption{Overlap reduction functions of the correlation between interferometers in cluster A and B (left panels) and cluster B and C (right panels) in the case I. The orientations of the interferometers are selected as $\sigma_A=\sigma_B=\sigma_C=0$. Solid (red), dotted (green), and dashed (blue) curves correspond to the tensor, vector, and scalar modes, respectively.}
\label{fig8}
\end{center}
\end{figure*}

In the four-cluster configuration,  
the detector separation $D$ is typically of the order of 1 AU. 
This means that 
the overlap reduction function starts to oscillate, 
and rapidly decay above the characteristic frequency, 
$f_c \equiv c/(2 D) \sim 10^{-3}\,\rm{Hz}$. Examples 
of the overlap reduction function for each polarization mode are 
shown in Fig.~\ref{fig8}, where the parameters of detector configuration 
are selected as $\sigma_A=\sigma_B=\sigma_C=0$, and the results for 
$\beta=30^{\circ}, 90^{\circ}$, and $150^{\circ}$ are plotted from top to 
bottom panels. 
At the frequency $f\sim0.1$Hz, the amplitudes of overlap reduction 
function are significantly dropped, but they 
show different oscillatory behaviors for each polarization mode. The latter
property is essential for separately detecting the scalar, vector 
and tensor GWBs. 
Note that the distance between BC link becomes identical for top and bottom 
panels, and the overlap reduction functions for each polarization mode 
coincide with each other.

In the present setup,  
the condition for separate detection of each polarization mode 
can be understood more precisely from Eqs.~(\ref{eq27}) - (\ref{eq28}). 
Consider the close detectors with $\alpha \ll 1$ and $\beta<1$.  
The spherical Bessel functions are approximated as
\begin{equation}
j_n(\alpha) \approx \frac{\alpha^n}{(2n+1)!!} \;. \nonumber
\end{equation}
Further, we obtain $U_1(\beta) = 16+{\cal{O}}(\beta^2)$ and 
$U_2(\beta) = 32 \beta+{\cal{O}}(\beta^3)$. Then, 
terms including $j_2$ and $j_4$ become negligible, and we have
\begin{eqnarray}
\gamma_{AB}^M &\approx & j_0(\alpha) [ 1+{\cal{O}}(\beta) ] \cos \bigl[2(\sigma_A-\sigma_B) \bigr] \;, \nonumber \\
\gamma_{BC}^M &\approx & j_0(\alpha) [ 1+{\cal{O}}(\beta) ] \cos \bigl[ 2(\sigma_B-\sigma_C) \bigr] \;. \nonumber 
\end{eqnarray}
Clearly, the overlap reduction functions for all polarization modes become 
degenerate, and 
reduce to an identical form in the limit $\beta\to0$. 
Therefore, the $j_2$ and $j_4$ terms play a crucial role in breaking this degeneracy. These terms become comparable to the $j_0$ term 
only when $\alpha\gtrsim1$, leading to the condition $D > \lambda_g$. 
Hence, widely separated detectors are essential for separately 
measuring each polarization mode of GWB. However, this generally conflicts 
with the optimal detection of GWBs. The resultant 
sensitivity to each polarization mode is thus significantly 
reduced, as shown below.

\begin{figure}[t]
\begin{center}
\includegraphics[width=8cm]{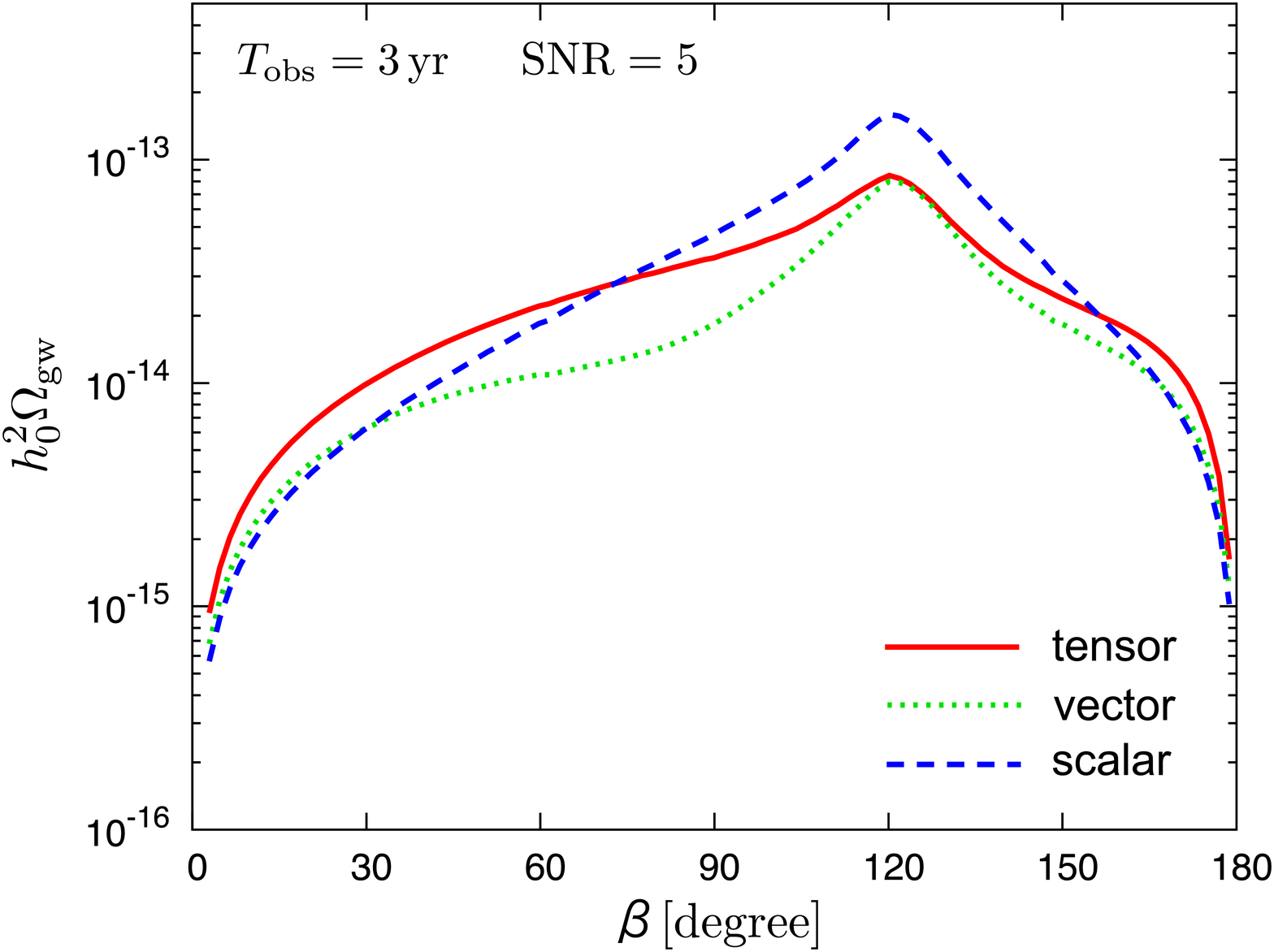}
\caption{Detectable $h_0^2 \Omega_{\rm{gw}}$ ($\xi h_0^2 \Omega_{\rm{gw}}$ for the scalar mode) after the mode separation in the case I.}
\label{fig2}
\end{center}
\end{figure}
\begin{figure}[t]
\begin{center}
\includegraphics[width=8cm]{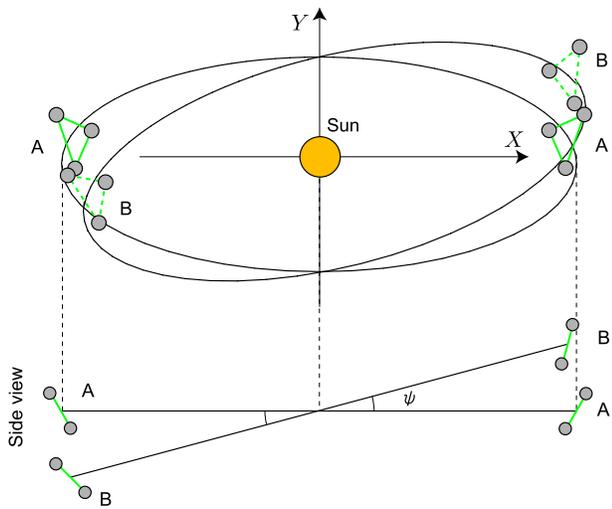}
\caption{Case II --- Two clusters, A and B, whose radius of the orbits is $1\,{\rm{AU}}$. The orbital plane of the cluster B is tilted by the angle $\psi$. }
\label{fig3}
\end{center}
\end{figure}

\subsubsection{Results}

The optimal SNR for four-cluster configuration is calculated with, in total, 
54 correlation signals (AA', AB, AC, A'B, A'C, BC 
$\times 9\; {\rm{links}}=54$). 
Setting the observation time and 
detection threshold to $T_{\rm{obs}}=3\,{\rm{yr}}$ and 
${\rm{SNR}}=5$, we estimate minimum detectable amplitude 
$h_0^2 \Omega_{\rm{gw}}$ ($\xi h_0^2 \Omega_{\rm{gw}}$ for the scalar mode).

In Fig.~\ref{fig2}, the resultant amplitude is plotted as function 
of angle $\beta$. At $\beta \sim 120^{\circ}$, the 
sensitivity degrades due to the symmetry of the detector configuration. 
In other words, the clusters, A, B, and C, are located at the apexes of an
equilateral triangle, and some of the correlation signals are degenerated. 
As $\beta$ approaches $0^{\circ}$ and $180^{\circ}$, the detector sensitivity 
reaches nearly maximum, because the clusters A and B (or C), and 
B and C are close to the colocated configuration. 
In practice, to keep a better angular resolution to point GW sources,  
two of four clusters have to be located 
far from the star-like clusters \cite{bib24,bib21}. 
Thus, the optimal choice of the parameter $\beta$ may be around 
$\beta\sim60^{\circ}$, which gives the detectable amplitude for each 
polarization mode as 
\begin{align}
&h_0^2 \Omega_{\rm{gw}}^T=2.2\times 10^{-14},
\nonumber\\
&h_0^2 \Omega_{\rm{gw}}^V=1.1\times 10^{-14},
\nonumber\\
&\xi h_0^2 \Omega_{\rm{gw}}^S=1.9\times 10^{-14}.
\nonumber
\end{align}
These results are compared with the optimal detection of GWB 
without mode separation. For two clusters that are colocated and 
coaligned like clusters A and A', the sensitivity reaches
$h_0^2 \Omega_{\rm{gw}}|_0 = 7.1 \times 10^{-17}$. Thus, for 
a separate detection of polarization modes, 
the sensitivity to GWB is significantly degraded by more than 
two orders of magnitude.

\subsection{Case II: two clusters}
\label{subsec:case2}

\subsubsection{Configuration}

For better measurement of each polarization of GWB,  
we consider an alternative setup shown in Fig.~\ref{fig3} originally proposed by Ref.~\cite{bib18} in order to detect 
a circularly polarized component of tensor GWB. 
In this setup, the orbital configuration of the one cluster 
is the same as cluster A in case I, while the orbital plane of 
the other cluster B is slightly tilted by the angle $\psi$ around the 
$Y$ axis. The position of each spacecraft in cluster A 
is described by Eqs.~(\ref{eq37}) and (\ref{eq38}), 
with specific choice of the parameters, $\theta=\pi/3$, $\sigma_1=0$, 
$\sigma_2=2\pi/3$, and $\sigma_3=4\pi/3$. The orbit of the guiding center 
of the cluster B, and the relative positions of the spacecrafts are 
respectively given by 
\begin{eqnarray}
\vec{X}_{B} (t) &=& \mathbf{R}_Y \bigl[ -\psi \bigr]\, \vec{X}_{A} (t) \;, 
\nonumber \\
\vec{\mathbf{x}}_{B} (t) &=& \mathbf{R}_Y \bigl[ -\psi \bigr]\, 
\vec{\mathbf{x}}_{A}(t) \;. 
\nonumber 
\end{eqnarray}

Note that, seen at a certain moment $t$, the intercluster correlation signals (the overlap reductions) between cluster
A and B are highly degenerate due to the geometrical degeneracy in
the interferometer location (e.g. $\gamma^T_{A1B1} \neq \gamma^T_{A1B2} \neq \gamma^T_{A1B3}$). So, the differences between the overlap reduction 
functions are of the order of ${\cal{O}}(L^2/D^2)$. 
However, the degeneracy can be broken by utilizing the 
orbital motion of the clusters. The advantage of this configuration is 
that the distance between the clusters gradually changes with time, 
and the correlation signals measured at the different times can be regarded as that of a different detector pair with different 
location and separation. As a result of closer detector separation, the 
sensitivity to each polarization mode can become even better 
compared to the four-cluster configuration.

As long as we consider the low-frequency GWs, 
the detector separation is approximately described by 
$D(\psi,\phi) = 2 R_0 \left| \sin (\psi/2)\,\cos \phi  \right|$. Since $\phi$ assigns a time to an overlap reduction function, the 
inclination angle $\psi$ is the only free parameter. In what follows, 
instead of $\psi$, we use the maximum separation of 
clusters, $D_{\rm{max}}=2R_0 |\sin (\psi/2)|$, to characterize 
the results.  

\subsubsection{Overlap reduction function}

Compared to the case I, 
the analytical expressions of overlap reduction functions for 
two-cluster configuration become much 
more complicated, but can be obtained from 
Eq. (\ref{eq25}). For a cross-correlation signal 
of AB links, the overlap reduction functions are
\begin{widetext}
\begin{eqnarray}
\gamma^M (f,\phi,\psi,\sigma_A,\sigma_B; \alpha) &=& \frac{1}{16} \biggl[ \Theta_1^M (\alpha,\phi,\psi)\, \cos[2(\sigma_A+\sigma_B)] + \Theta_2^M (\alpha, \phi,\psi)\, \sin [2(\sigma_A+\sigma_B) \nonumber \\
&& \quad +\Theta_3^M (\alpha,\phi,\psi)\, \cos[2(\sigma_A-\sigma_B)] + \Theta_4^M (\alpha, \phi,\psi)\, \sin [2(\sigma_A-\sigma_B)] \biggr] \;, \nonumber
\end{eqnarray}
where
\begin{eqnarray} 
\left(
\begin{array}{c} 
\Theta_{1,2}^T \\ \Theta_{1,2}^V \\ \Theta_{1,2}^S 
\end{array}
\right)
& \equiv & - \left(
\begin{array}{c} 
J_T \\ J_V \\ J_S 
\end{array}
\right) U_{1,2}(\phi) \sin^4 \left( \frac{\psi}{2} \right) 
\pm \frac{45}{56} \left(
\begin{array}{c} 
- 8j_2- j_4 \\ 4 j_2 + 4 j_4 \\ 8 j_2 -6 j_4 
\end{array}
\right) V_{1,2}(\phi) \sin^2 \left( \frac{\psi}{2} \right)
-\frac{45}{16} \left(
\begin{array}{c} 
j_4 \\ -4 j_4 \\ 6 j_4 
\end{array} \right)
\cos 4\phi \;, \nonumber \\
&& \label{eq31} \\
\left(
\begin{array}{c} 
\Theta_3^T \\ \Theta_3^V \\ \Theta_3^S  
\end{array}
\right)
& \equiv & \left(
\begin{array}{c} 
J_T \vspace{.5em} \\ J_V \vspace{.5em} \\ J_S
\end{array}
\right) U_3 (\phi) \sin^4 \left( \frac{\psi}{2} \right)
-8 
\left[ 4 
\left( 
\begin{array}{c}
j_0 +\frac{25}{56} j_2 - \frac{3}{448} j_4 \vspace{.5em} \\ j_0 -\frac{25}{112} j_2 + \frac{3}{112} j_4 \vspace{.5em} \\ j_0 -\frac{25}{56} j_2 - \frac{9}{224} j_4
\end{array} 
\right) \right. \nonumber \\
&& \left. +9 \left( 
\begin{array}{c}
j_0 +\frac{5}{8} j_2 + \frac{1}{64} j_4 \vspace{.5em} \\ j_0 -\frac{5}{16} j_2 - \frac{1}{16} j_4 \vspace{.5em} \\ j_0 -\frac{5}{8} j_2 + \frac{3}{32} j_4  
\end{array} 
\right) \sin^2 \phi 
\right] \sin^2 \left( \frac{\psi}{2} \right)
+ 16 \left( 
\begin{array}{c}
j_0 +\frac{5}{28} j_2 - \frac{37}{1792} j_4 \vspace{.5em} \\ j_0 -\frac{5}{56} j_2 + \frac{37}{448} j_4 \vspace{.5em} \\ j_0 -\frac{5}{28} j_2  - \frac{111}{896} j_4 
\end{array} 
\right) \;,  \\
\left(
\begin{array}{c} 
\Theta_4^T \\ \Theta_4^V \\ \Theta_4^S  
\end{array}
\right)
& \equiv & -16 \sqrt{3} \sin \psi \sin \phi 
\left[ \left(
\begin{array}{c} 
J_T \vspace{.5em} \\ J_V \vspace{.5em} \\ J_S 
\end{array}
\right) \biggl( 1+\frac{3}{4} \sin^2 \phi \biggr) \sin^2 \left( \frac{\psi}{2} \right) 
- \left( 
\begin{array}{c}
j_0 +\frac{25}{56} j_2 - \frac{3}{448} j_4 \vspace{.5em} \\ j_0-\frac{25}{112} j_2 + \frac{3}{112} j_4 \vspace{.5em} \\ j_0 -\frac{25}{56} j_2 - \frac{9}{224} j_4 
\end{array} 
\right) \right] \;, 
\label{eq32} 
\end{eqnarray}
\end{widetext} 
\begin{eqnarray}
U_3(\phi) &=& 97-90 \cos^2 \phi +9\cos^4 \phi \;, \nonumber \\
V_1(\phi) &=& 1+3\cos^2 \phi -8 \cos^6 \phi \;, \nonumber \\
V_2(\phi) &=& 2\sin 2\phi \cos^2 \phi (1+2 \cos^2 \phi ) \;. \nonumber 
\end{eqnarray}
The functions, $U_1$, $U_2$, $J_T$, $J_V$, and $J_S$, are defined 
in Eqs. (\ref{eq29})-(\ref{eq30}), and the dimensionless frequency 
$\alpha$ is defined by 
Eq. (\ref{eq26}). The argument of the spherical Bessel functions 
and $\Theta_{1,2,3,4}$ are omitted in the above equations. 
Note also that the dimensionless quantity $\alpha$ depends on 
not only $\phi$ but also $\psi$. 
\begin{figure*}[t]
\begin{center}
\includegraphics[width=15cm]{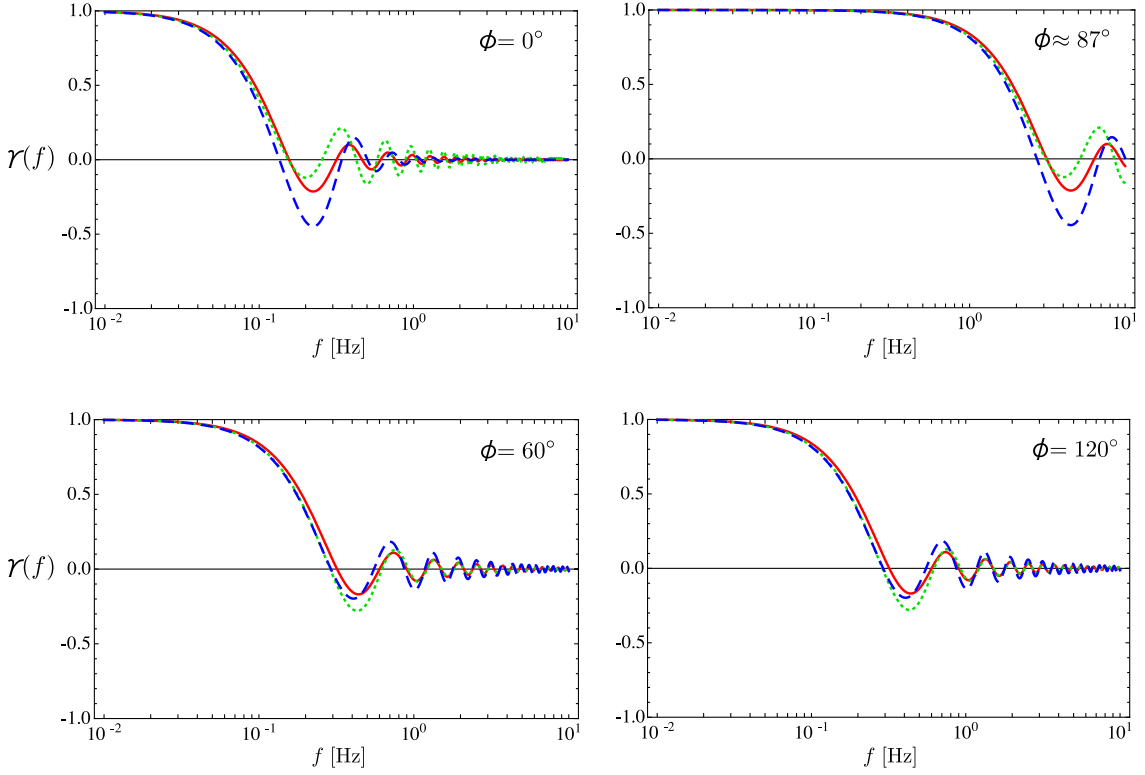}
\caption{Overlap reduction functions of the correlation between interferometers in cluster A and B (AB link) in the case II. The orientations of the interferometers are selected as $\sigma_A=\sigma_B=0$. The tilt of the orbit is fixed to $D_{\rm{max}}/L =10^3$. Solid (red), dotted (green), and dashed (blue) curves correspond to the tensor, vector, and scalar modes, respectively.}
\label{fig9}
\end{center}
\end{figure*}

The examples of the overlap reduction function for each polarization 
mode are shown in Fig. \ref{fig9}, where the 
parameters of the detector configuration are specifically chosen as 
$\sigma_A=\sigma_B=0$ and $D_{\rm{max}}/L =10^3$,  
and the results are shown for $\phi=0^{\circ}$ (top left), 
$60^{\circ}$ (bottom left), $87^{\circ}$ (top right), and 
$120^{\circ}$ (bottom right). 
The distance between clusters A and B becomes maximum at 
$\phi=0^{\circ}$ and $180^{\circ}$, and is minimum at $\phi=90^{\circ}$ 
and $270^{\circ}$. Compared to the four-cluster case, 
the amplitudes of overlap reductions functions at $f\sim 0.1$ Hz  
become even larger for each polarization mode. This implies that 
the separate detection of polarized GWBs is achievable with high SNR.

As examined in the four-cluster configuration,  we consider the 
condition for a separate detection of three polarization modes. 
For small $\psi \ll 1$, Eqs.~(\ref{eq31})-(\ref{eq32}) become 
\begin{eqnarray} 
\left(
\begin{array}{c} 
\Theta_{1,2}^T \\ \Theta_{1,2}^V \\ \Theta_{1,2}^S 
\end{array}
\right)
& = & -\frac{45}{16} \left(
\begin{array}{c} 
j_4 \\ -4 j_4 \\ 6 j_4 
\end{array} \right)
\cos 4\phi \;, \nonumber \\
\left(
\begin{array}{c} 
\Theta_3^T \\ \Theta_3^V \\ \Theta_3^S 
\end{array}
\right)
& = & 16 \left( 
\begin{array}{c}
j_0 +\frac{5}{28} j_2 - \frac{37}{1792} j_4 \vspace{.5em} \\ j_0 -\frac{5}{56} j_2 + \frac{37}{448} j_4 \vspace{.5em} \\ j_0 -\frac{5}{28} j_2  - \frac{111}{896} j_4 
\end{array} 
\right) \;,  \nonumber \\
\left(
\begin{array}{c} 
\Theta_4^T \\ \Theta_4^V \\ \Theta_4^S  
\end{array}
\right) 
& = & 16 \sqrt{3}\, \psi \sin \phi \nonumber \\
&& \times \left( 
\begin{array}{c}
j_0 +\frac{25}{56} j_2 - \frac{3}{448} j_4 \vspace{.5em} \\ j_0-\frac{25}{112} j_2 + \frac{3}{112} j_4 \vspace{.5em} \\ j_0 -\frac{25}{56} j_2 - \frac{9}{224} j_4 
\end{array} 
\right) \;. \nonumber
\end{eqnarray}
Thus, for $\alpha\ll1$, the terms $j_2$ and $j_4$ become negligible,  
and the overlap reduction functions for all polarization modes are reduced 
to an identical form. 
Hence, $j_0\sim j_2\sim j_4$ is required for 
having the different oscillatory behaviors for overlap reduction function of 
scalar, vector, and tensor modes and leads to the same 
conclusion as that in case I configuration, $D > \lambda_g$ for each pair of detectors.

\begin{figure}[h]
\begin{center}
\includegraphics[width=8cm]{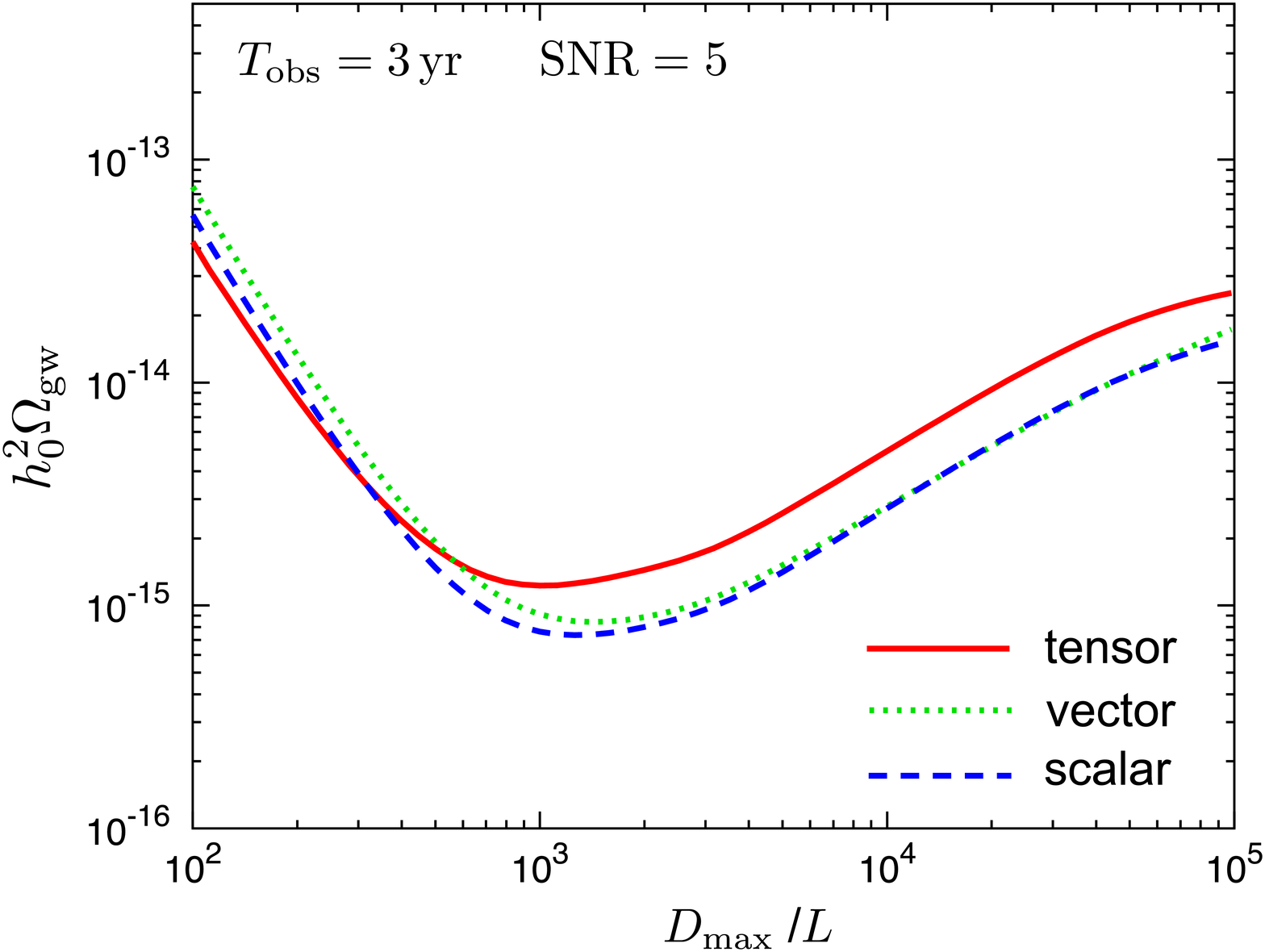}
\caption{Detectable $h_0^2 \Omega_{\rm{gw}}$ ($\xi h_0^2 \Omega_{\rm{gw}}$ for the scalar mode) after the mode separation in the case II.}
\label{fig4}
\end{center}
\end{figure}

\subsubsection{Results}

For two-cluster configuration, 
we have, in total, 9 correlation signals between clusters A and B. 
Combining these signals, the optimal SNR is computed taking account 
of the time variation of the overlap reduction functions. 
In practice, the time 
integral in Eq. (\ref{eq3}) is discretized as the sum of the 
finite time segment. We checked that the results remain unchanged 
if the number of segments in one year is larger than twelve.

In Fig.~\ref{fig4}, minimum detectable amplitude $h_0^2 \Omega_{\rm{gw}}$ 
is plotted against the maximum separation, $D_{\rm{max}}/L$, keeping 
$T_{\rm{obs}}=3\,{\rm{yr}}$, and ${\rm{SNR}}=5$. The detectable amplitude 
for each polarization mode first decreases and begins increasing as the separation $D_{\rm{max}}$ increases. The best sensitivity is 
achieved at $D_{\rm{max}}/L\sim1.4\times 10^3$, and the detectable amplitude 
for each polarization mode becomes
\begin{align}
&h_0^2 \Omega_{\rm{gw}}^T=1.3\times 10^{-15}, 
\nonumber\\
&h_0^2 \Omega_{\rm{gw}}^V=8.5\times 10^{-16}, 
\nonumber\\
&\xi h_0^2 \Omega_{\rm{gw}}^S=7.4\times 10^{-16}. 
\nonumber
\end{align}
Therefore, compared to the four-cluster configuration, 
the sensitivity to the separate detection of each polarization mode 
is greatly improved. Note, however, that 
the optimal sensitivity to GWBs themselves is rather degraded, compared with those when we do not consider the mode separation. 
This is because no colocated and coaligned cluster 
configuration are available in the present setup.

\section{Discussion}
\label{sec:discussion}

The previous section reveals that a separate detection of three polarization 
modes with high signal sensitivity needs a sophisticated setup for detector 
configuration, but we may achieve $\Omega_{\rm gw}\sim10^{-15}$. In this 
section, we briefly discuss how the results are changed for a different 
setup or situation.

First consider the influence of astrophysical foregrounds, which  
was not taken into account when we estimated the SNR. It is 
expected that the low-frequency side of the DECIGO could be dominated 
by the unresolved GWs from the white-dwarf binaries. 
According to the estimation by Ref.~\cite{bib8}, cosmological population 
of white-dwarf binaries produces a large GW signal at 
$f\lesssim0.2$Hz, and may act as a confusion noise.  
Thus, below the frequency $f_{\rm cut}=0.2$Hz, a definite detection of 
cosmological GWBs might not be possible.

\begin{figure}[t]
\begin{center}
\includegraphics[width=8cm]{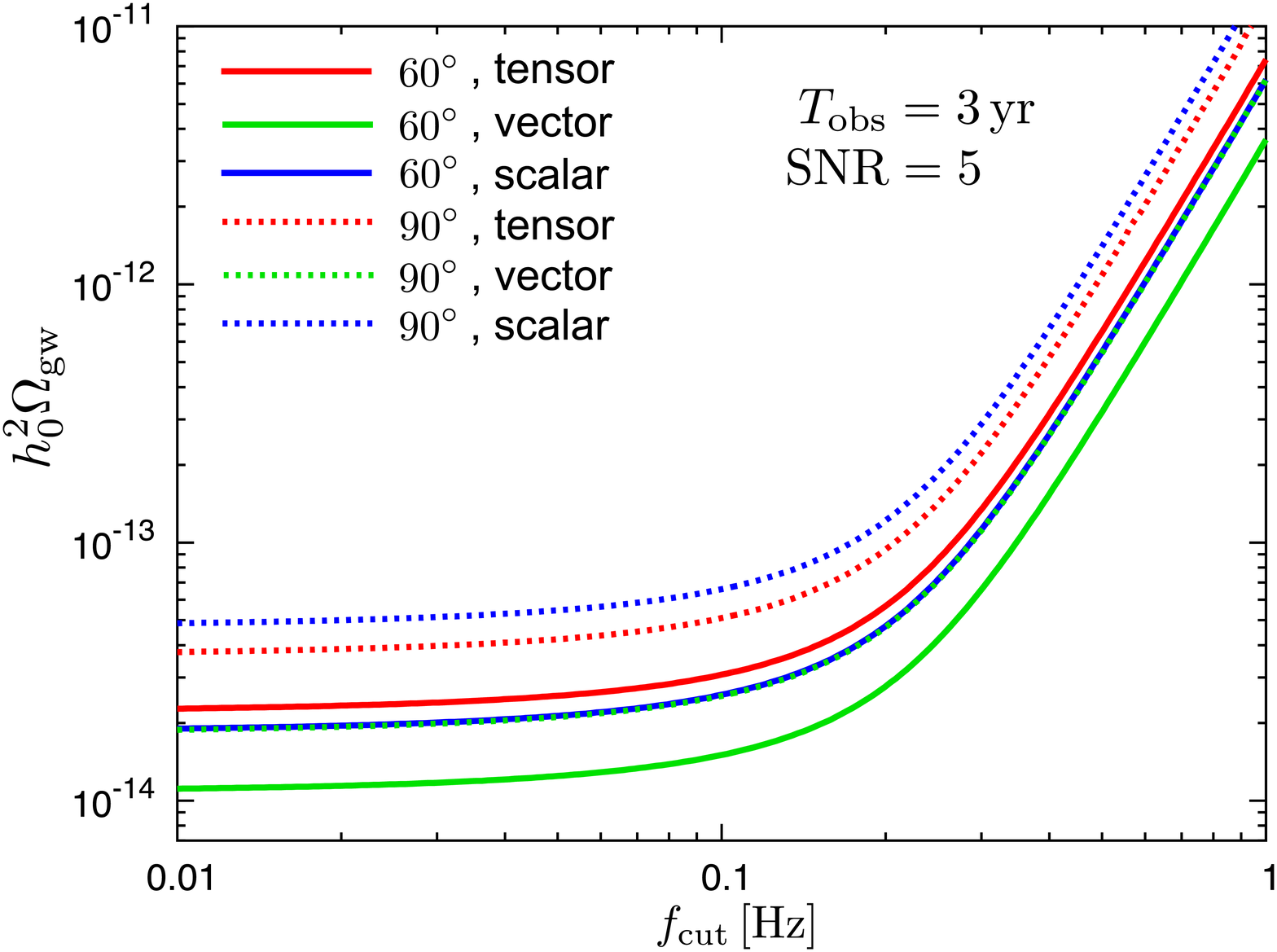}
\caption{Detectable $h_0^2 \Omega_{\rm{gw}}$ ($\xi h_0^2 \Omega_{\rm{gw}}$ for the scalar mode) after the mode separation in the case I with the cutoff frequency. The detector separation is selected as $\beta=60^{\circ}$ (solid curves) and $\beta=90^{\circ}$ (dotted curves).}
\label{fig6}
\end{center}
\end{figure}

\begin{figure}[t]
\begin{center}
\includegraphics[width=8cm]{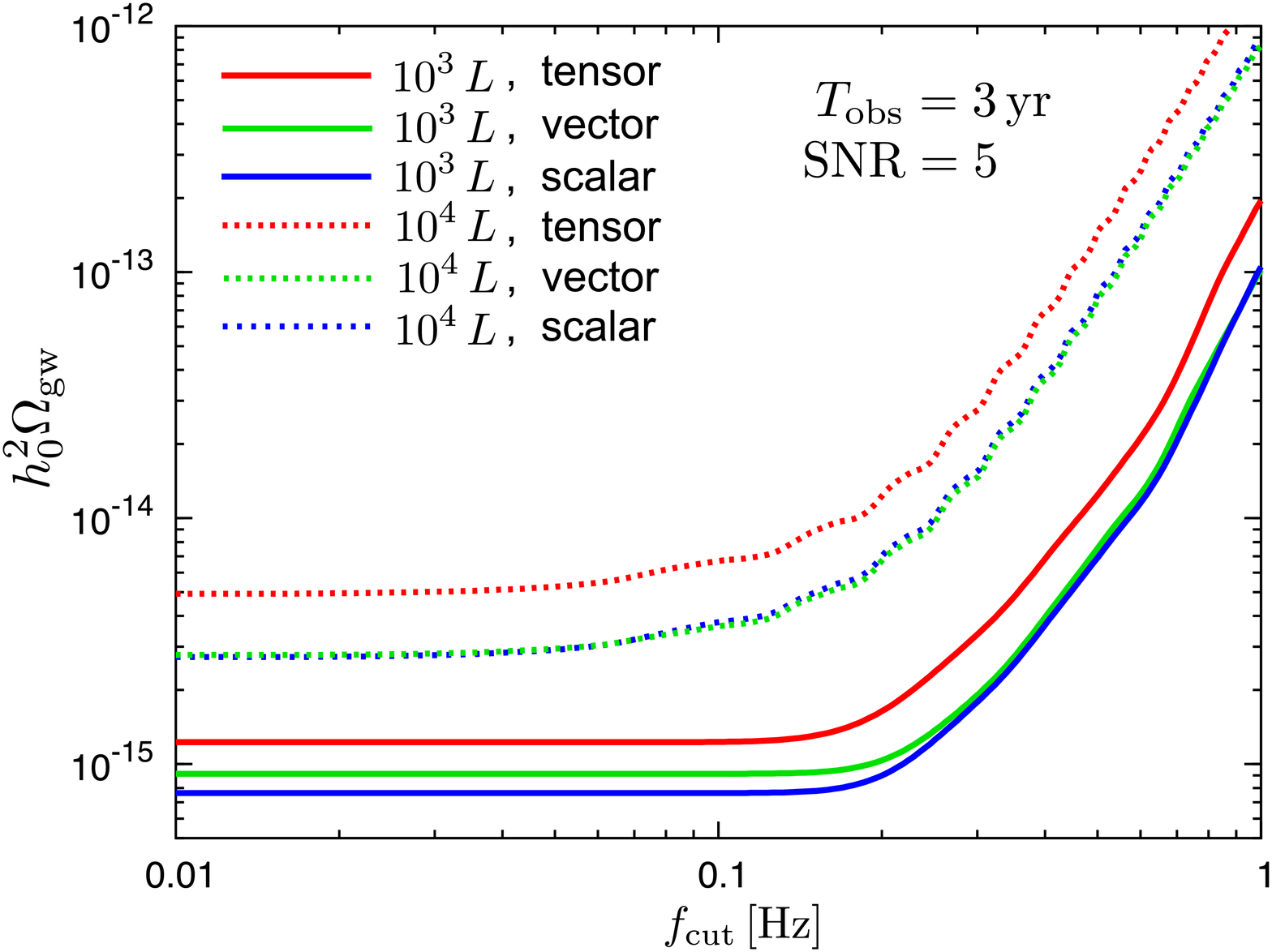}
\caption{Detectable $h_0^2 \Omega_{\rm{gw}}$ ($\xi h_0^2 \Omega_{\rm{gw}}$ for the scalar mode) after the mode separation in the case II with the cutoff frequency. The detector separation is selected as $D_{\rm{max}}=10^3 L$ (solid curves) and $10^4 L$ (dotted curves).}
\label{fig7}
\end{center}
\end{figure}

Here, in order to examine the significance of this effect, we 
introduce the low-frequency cutoff in the frequency integral 
of Eq.~(\ref{eq2}), and estimate the SNR again. Based on this, 
the detectable amplitude of GWB is calculated for four- and two-cluster 
configurations. In Figs.~\ref{fig6} and \ref{fig7}, the dependence of 
the detectable energy density $h_0^2 \Omega_{\rm{gw}}$ on 
the cutoff frequency are shown for 
four-cluster case with $\beta=60^{\circ}$ and $90^{\circ}$, and 
two-cluster setup with $D_{\rm{max}}/L=10^3$ and $10^4$, respectively.  
In both cases, the effect of low-frequency cutoff becomes significant as $f_{\rm cut}$ increases, but the results are not drastically 
changed at $f_{\rm cut}\lesssim 0.2$Hz. This is rather consistent with the 
results by Ref.~\cite{bib66}. Thus, even in the presence of 
confusion noise, the detectable $h_0^2 \Omega_{\rm{gw}}$ remains unchanged 
as long as the cutoff frequency is below $0.2\,{\rm{Hz}}$.  
This conclusion may be rather natural because DECIGO 
has been designed to evade the low-frequency confusion noises.  


Next consider the alternative design of space interferometer, i.e., BBO. 
As we mentioned in Sec.~\ref{sec:intro}, 
BBO plans to use a transponder type with the arm length, 
$L=5\times10^4\,\rm{km}$. This point is rather different from 
DECIGO, however, the noise curve and the detector 
configuration of BBO are almost the same as those of DECIGO. Thus, 
we naively expect that the results obtained in the previous section 
basically hold for the case of BBO. A subtle point is that 
low-frequency approximation of the detector response which we adopted 
throughout the analysis might not be valid for 
GWs at frequencies $f\gtrsim1$Hz. Thus, a correct treatment without 
using low-frequency approximation is necessary for quantitative 
estimation of detectability. Nevertheless, quantitative difference 
would be certainly small, and the qualitative point of our results 
can be applied to the BBO case, because the noise curve of BBO coincides with that of DECIGO within $5\,\%$ of a factor at frequencies below $1\,{\rm{Hz}}$ \cite{bib21} and the SNR is almost determined by a signal below $1\,{\rm{Hz}}$. 

\section{Summary}
\label{sec:summary}

In this paper, we discuss how well we can separately detect and measure 
the extra-polarization modes of a GWB in addition to the 
standard tensor-type GWB, i.e., scalar and vector GWBs,  
via the space-based interferometers. In addition to the tensor mode, 
scalar and vector-type GWBs may have been produced in the early 
stage of the Universe through various mechanisms including inflation, 
phase transition and reheating of the Universe, when the general 
relativity would not strictly hold. Thus, the detection and measurement of 
scalar and vector modes of GWBs is a direct probe of gravity, and can also yield information about the physics of the early universe.

We have presented the formula for optimal SNR to separately detect 
three polarization modes combining multiple correlation signals. 
Based on this, we have considered the two specific configurations for 
planned space interferometer, DECIGO, and estimated the ability to 
detect extra-polarization modes of GWB. For the four-cluster setup 
consisting of the four sets of spacecraft constellations with coplanar 
orbits, the detectable minimum amplitudes of GWB are degraded 
significantly for each polarization mode, and the detectable density 
somehow reaches $h_0^2 \Omega_{\rm{gw}}\sim10^{-14}$. This is in 
marked contrast to the standard analysis that only considers the tensor 
mode of GWB. To raise the sensitivity, we then considered the two-cluster 
setup, in which the orbits of two set of spacecraft constellation are 
slightly misaligned. Thanks to the non-stationarity of detector 
configuration, the cross-correlation measured at different times can be 
regarded as an independent set of signals with different location and 
separation, and this helps to improve the detection sensitivity. As a 
result, the detectable density is found to be 
$h_0^2 \Omega_{\rm{gw}}\sim 10^{-15}$ for tensor mode, and 
even better for scalar and vector modes.

Currently, no definite theoretical bound on the amplitude of GWB exists 
below $h_0^2 \Omega_{\rm{gw}}\sim 10^{-6}$, and thus 
it is rather difficult to predict how much amount of GWB is expected 
for each polarization mode. Nevertheless, constraints from
cosmic microwave background anisotropies imply that the 
tensor-type GWB generated during inflation is likely to be as small as 
$h_0^2 \Omega_{\rm{gw}} \sim 10^{-16}$ at frequency $f\sim0.1$-$1\,{\rm{Hz}}$ \cite{bib54,bib55,bib56,bib61}.
Given that the coupling parameters of scalar 
and vector degrees of freedom to a background gravitational field are 
almost the same as that of the tensor, no significant amount of GWB is 
expected for scalar and vector polarizations. Therefore, with the 
setup examined in this paper, it might be hard to separately 
measure the three polarization modes of inflationary GWB, although  
the detector itself has an ability to detect such a small GWB.

Nonetheless, this argument is based on an extrapolation 
from the extremely low-frequency observation by 16 orders of magnitude, 
and there may still exist many windows to generate a large 
amplitude of inflationary GWB around frequency $f\sim0.1$-$1\,{\rm{Hz}}$. 
Further, there are several viable scenarios that can produce 
a large amplitude of low-frequency GWB. An example is the GWB produced 
by density fluctuation through cosmological phase transition and/or 
preheating. In this case, the energy density of scalar GWB might exceed 
that of the tensor mode, because scalar GW would be easily emitted by the 
monopole moment of the density fluctuation. The resultant spectrum of GWB 
may has a sharp peak with the amplitude of at most 
$h_0^2 \Omega_{\rm{gw}} \sim 10^{-7}$ \cite{bib47,bib50,bib51,bib53,bib58}. 
Hence, even with a limited sensitivity, a search for additional polarization 
modes of GWB via space-based interferometer is indispensable for 
a cosmological test of gravity, 
and definitely yields an additional scientific benefit for probing the 
physics of the early universe.

\begin{acknowledgments}
We thank N. Kanda, M. Sakagami and N. Seto for helpful comments and 
discussions. A. T. is supported in part by a Grants-in-Aid for Scientific
Research from the JSPS under Grant No. 21740168.
\end{acknowledgments}
\appendix

\section{Correlation signal in a cluster}
\label{appB}

In this Appendix, 
we show that it is impossible to obtain a correlation signal sensitive to a GWB in a cluster, even if three interferometers in a cluster are used.

Let us consider a correlation signal in a cluster of DECIGO like Fig. \ref{fig10}. We denote three spacecrafts as SC1, SC2, and SC3, three interferometers in the cluster as IFO1, IFO2, and IFO3. The displacement noise of the optical link between i-th SC and j-th SC (the light is injected from i-th SC, reflected at the mirror near j-th SC, and finally returns to i-th SC.) is $d_{ij}$ and the shot noise at i-th SC is $\zeta_i$. The noise components in the signal obtained by each IFO are written as 
\begin{eqnarray}
s_1&=& d_{12}-d_{13}+\zeta_1 \;, \label{eq33} \\
s_2&=& d_{23}-d_{21}+\zeta_2 \;, \\
s_3&=& d_{31}-d_{32}+\zeta_3 \;. \label{eq34} 
\end{eqnarray}
The displacement noises, $d_{ij}$, can be considered to be symmetric with respect to the subscripts, because the cavity storage time of light, $10L/c \approx 0.03\,{\rm{sec}}$, is shorter than the period of a GW that we are interested in, $\sim 1-10\,{\rm{sec}}$. So, Eqs. (\ref{eq33}) - (\ref{eq34}) can be written as 
\begin{eqnarray}
s_1&=& d_{12}-d_{13}+\zeta_1 \;, \label{eq35} \\
s_2&=& d_{23}-d_{12}+\zeta_2 \;,  \\
s_3&=& d_{13}-d_{23}+\zeta_3 \;. \label{eq36}
\end{eqnarray}
We linearly combine Eqs. (\ref{eq35}) - (\ref{eq36}) with arbitrary coefficients, and take the ensemble average of the correlation signal 
\begin{eqnarray}
&&\langle (s_1+c_2 s_2+c_3 s_3)(s_1+c_2^{\prime} s_2+c_3^{\prime} s_3) \rangle \nonumber \\
&&= (1-c_2)(1-c_2^{\prime}) \langle d_{12}^2 \rangle + (1-c_3)(1-c_3^{\prime}) \langle d_{13}^2 \rangle   \nonumber \\
&&  + (c_2-c_3)(c_2^{\prime}-c_3^{\prime}) \langle d_{23}^2 \rangle + ( 1 + c_2 c_2^{\prime} + c_3 c_3^{\prime} )\langle \zeta^2 \rangle \;. \nonumber \\
&&
\end{eqnarray}
Here we assumed that $\langle d_{ij} \zeta_k \rangle =0$, $\langle d_{ij} d_{k\ell} \rangle =0$ for $i \neq j$ and $k \neq \ell$, and $\langle \zeta_i \zeta_j \rangle =0$ for $i \neq j$, and wrote $\langle \zeta^2 \rangle = \langle \zeta_i \zeta_j \rangle$ for $i=j$. To obtain the correlation signal that is insensitive to the correlation noise, the coefficients should be chosen as
\begin{equation}
(c_2, c_3, c_2^{\prime}, c_3^{\prime}) = (c_2, -1-c_2,1,1)\;,  \nonumber
\end{equation} 
where $c_2$ is arbitrary. Thus one of the combination signals in the correlation must be a symmetric combination, $s_1+s_2+s_3$, regardless of another combination signal. 

Next, we define 
\begin{equation}
s_{\rm{sym}} \equiv \frac{1}{\sqrt{3}} (s_1+s_2+s_3) \;, \nonumber
\end{equation}
which is known as the symmetrized Sagnac signal \cite{bib33,bib34,bib35}, and calculate its GW signal. As we will see below, the correlation signal with $s_{\rm{sym}}$ is not helpful for the separation of the multiple polarization modes. Using Eq. (\ref{eq7}), we can find the GW signal of $s_{\rm{sym}}$
\begin{eqnarray}
h_{\rm{sym}} &\equiv & \frac{1}{\sqrt{3}} (h_1+h_2+h_3) \nonumber \\
&=& \sum_p \int _{S^2} d\hat{\mathbf{\Omega}} \int_{-\infty}^{\infty}df\, \nonumber \\
&& \times \tilde{h}_p (f, \hat{\mathbf{\Omega}})\, e^{2\pi i f t}\, \mathbf{D}_{\rm{sym}} : \mathbf{e}_p (\hat{\mathbf{\Omega}}) \:, \nonumber 
\end{eqnarray}
where $\mathbf{D}_{\rm{sym}}$ at the zeroth order in the low frequency approximation is exactly zero since all terms are canceled due to the symmetry of the combination. To obtain the leading contribution, one needs to include the response functions in the detector arms in Eq. (\ref{eq1}) \cite{bib30,bib33,bib34,bib31,bib32}. The detector tensor of the combination signal $s_{\rm{sym}}$ is $\mathbf{D}_{\rm{sym}} (f, \hat{\mathbf{\Omega}}) \propto i f L/c$, then the GW response is suppressed below the frequency, $f \approx c/L \approx 300 \,{\rm{Hz}}$. Consequently, at $0.1-1\,{\rm{Hz}}$, the GW response is $300-3000$ times worse than that before taking the signal combination.

\section{Derivation of optimal SNR formula for separately detecting 
scalar, vector and tensor polarizations}
\label{appA}

Here, we will derive the SNR formula for separately detecting 
the three polarization modes by optimally combining arbitrary number 
of detector signals ($N_{\rm{pair}} \geq 3$). 

When one correlates detector signals in a frequency bin, the estimated value of the correlation signal, $\hat{\mu}_i(f)$, fluctuates around the true value, $\mu_i(f)$. Assuming the width of a frequency bin is much larger than the frequency resolution of the data we obtain, the likelihood function for $\hat{\mu}_i(f)$ is expected to be Gaussian distribution, owing to the central limit theorem. Let us denote a set of the estimated correlation signals of a detector pair in a frequency bin, $\{ \hat{\mu}_i(f); 1 \leq i \leq N_{\rm{pair}} \}$, where the subscript $i$ designates a detector pair (for I-th and J-th detector pair, $i=IJ$). The multidimensional likelihood function for the set of the estimator is written as
\begin{equation}
L\bigl[ \{ \hat{\mu}_i(f) \} \bigr] \propto \exp \left[ -\sum_{i=1}^{N_{\rm{pair}}} \frac{\{ \hat{\mu}_i(f)-\mu_i(f) \}^2}{{2 \cal{N}}_i(f)} \right] \;,
\label{eq39}
\end{equation}
where the covariance noise matrix, ${\cal{N}}_{i}(f)$, is defined as, say, ${\cal{N}}_{12}(f) \equiv P_1(f) P_2(f)$. Note that we assume that detector noise is not correlated with other detectors and that a GW signal is much smaller than the noise, so the calculation in Eq. \ref{eq24} is applied here. 

On the other hand, from Eq. (\ref{eq23}), the GW contribution in the correlation signal is
\begin{eqnarray}
\mu_i(f) &\propto & \gamma_i^T (f) \Omega_{\rm{gw}}^T(f) + \gamma_i^V (f) \Omega_{\rm{gw}}^V(f) + \gamma_i^S (f) \xi \Omega_{\rm{gw}}^S(f) \;, \nonumber \\
&& \label{eq40} \\
\hat{\mu}_i(f) &\propto & \gamma_i^T (f) \hat{\Omega}_{\rm{gw}}^T(f) + \gamma_i^V (f) \hat{\Omega}_{\rm{gw}}^V(f) + \gamma_i^S (f) \xi \hat{\Omega}_{\rm{gw}}^S(f) \;. \nonumber \\
&& \label{eq41}
\end{eqnarray}
The hat fixed to $\Omega_{\rm{gw}}$ represents that it is the estimated quantity. Substituting Eqs. (\ref{eq40}) and (\ref{eq41}) for Eq. (\ref{eq39}), we obtain the quadratic with respect to $\Omega_{\rm{gw}}$ in the argument of Eq. (\ref{eq39}), from which we can read the proportional relation of the Fisher matrix 
\begin{equation}
\mathbf{F} (f) = {\rm{a\;factor}} \times \left(
\begin{array}{ccc} 
F_{TT} & F_{TV} & F_{TS} \\
F_{TV} & F_{VV} & F_{VS} \\
F_{TS} & F_{VS} & F_{SS}
\end{array}
\right) \;, \nonumber
\end{equation}
with
\begin{eqnarray}
F_{TT}(f) &=& \sum_i \frac{(\gamma_i^T)^2}{{\cal{N}}_i} \;, \quad F_{VV}(f) = \sum_i \frac{(\gamma_i^V)^2}{{\cal{N}}_i} \;, \quad \label{eq45} \\
F_{SS}(f) &=& \sum_i \frac{(\gamma_i^S)^2}{{\cal{N}}_i} \;, \quad F_{TV}(f) = \sum_i \frac{\gamma_i^T \gamma_i^V}{{\cal{N}}_i} \;, \quad \\
F_{VS}(f) &=& \sum_i \frac{\gamma_i^V \gamma_i^S}{{\cal{N}}_i} \;, \quad
F_{TS}(f) = \sum_i \frac{\gamma_i^T \gamma_i^S}{{\cal{N}}_i} \;, \quad \label{eq46}
\end{eqnarray}
Thus, we find that the SNR in a frequency bin is proportional to some combination of the components of the Fisher matrix, namely
\begin{eqnarray}
\bigl[{\rm{SNR}}^T(f) \bigr]^2 & \propto & \frac{(\Omega_{\rm{gw}}^T)^2}{(\mathbf{F}^{-1})_{11}} = \frac{(\Omega_{\rm{gw}}^T)^2 \det \mathbf{F}}{F_{VV} F_{SS}- F_{VS}^2} \;, \label{eq42} \\
\bigl[{\rm{SNR}}^V(f) \bigr]^2 & \propto & \frac{(\Omega_{\rm{gw}}^V)^2}{(\mathbf{F}^{-1})_{22}} = \frac{(\Omega_{\rm{gw}}^V)^2 \det \mathbf{F}}{F_{TT} F_{SS}- F_{TS}^2} \;, \\
\bigl[{\rm{SNR}}^S(f) \bigr]^2 & \propto & \frac{(\xi \Omega_{\rm{gw}}^S)^2}{(\mathbf{F}^{-1})_{33}} = \frac{(\Omega_{\rm{gw}}^S)^2 \det \mathbf{F}}{F_{TT} F_{VV}- F_{TV}^2}\;. \label{eq43}
\end{eqnarray}

To determine the frequency-dependent factor of the proportional relation, we compare those with the SNR formula for $N_{\rm{pair}}=3$ case with three detectors, which has been derived in \cite{bib12}. For $N_{\rm{pair}}=3$, Eqs. (\ref{eq42}) - (\ref{eq43}) are reduced to
\begin{eqnarray}
({\rm{SNR}}^M(f))^2 &\propto & \left[ \frac{H_g^2(f)}{H_n^2(f)} \right] ^M \;, \quad M=T,V,S\;, \nonumber \\
&& \label{eq44}
\end{eqnarray}
\begin{equation}
\mathbf{\Pi}(f) \equiv
\left(
\begin{array}{ccc} 
\gamma^T_{12} & \gamma^V_{12} & \gamma^S_{12} \\
\gamma^T_{23} & \gamma^V_{23} & \gamma^S_{23} \\
\gamma^T_{31} & \gamma^V_{31} & \gamma^S_{31}
\end{array}
\right) \;, \nonumber
\end{equation}
\begin{widetext}
\begin{eqnarray}
\left[ \frac{H_g^2(f)}{H_n^2(f)} \right]^T &=& \frac{(\Omega_{\rm{gw}}^T)^2 (\det \mathbf{\Pi})^2}{P_1 P_2(\gamma_{23}^V \gamma_{31}^S - \gamma_{23}^S \gamma_{31}^V)^2+P_2 P_3(\gamma_{31}^V \gamma_{12}^S - \gamma_{31}^S \gamma_{12}^V)^2+P_3 P_1(\gamma_{12}^V \gamma_{23}^S - \gamma_{12}^S \gamma_{23}^V)^2} \;, \nonumber \\
\left[ \frac{H_g^2(f)}{H_n^2(f)} \right]^V &=& \frac{(\Omega_{\rm{gw}}^V)^2 (\det \mathbf{\Pi})^2}{P_1 P_2(\gamma_{23}^S \gamma_{31}^T - \gamma_{23}^T \gamma_{31}^S)^2+P_2 P_3(\gamma_{31}^S \gamma_{12}^T - \gamma_{31}^T \gamma_{12}^S)^2+P_3 P_1(\gamma_{12}^S \gamma_{23}^T - \gamma_{12}^T \gamma_{23}^S)^2} \;, \nonumber \\
\left[ \frac{H_g^2(f)}{H_n^2(f)} \right]^S &=& \frac{(\xi \Omega_{\rm{gw}}^S)^2 (\det \mathbf{\Pi})^2}{P_1 P_2(\gamma_{23}^T \gamma_{31}^V - \gamma_{23}^V \gamma_{31}^T)^2+P_2 P_3(\gamma_{31}^T \gamma_{12}^V - \gamma_{31}^V \gamma_{12}^T)^2+P_3 P_1(\gamma_{12}^T \gamma_{23}^V - \gamma_{12}^V \gamma_{23}^T)^2} \;. \nonumber
\end{eqnarray} 
\end{widetext}

On the other hand, according to \cite{bib12}, the SNR formula in $N_{\rm{pair}} = 3$ case (A factor coming from non-orthogonal arms, $\sin \chi^2=3/4$, is corrected.) is given by
\begin{equation}
{\rm{SNR}} = \frac{9 H_0^2}{40 \pi^2} \sqrt{T_{\rm{obs}}} \biggl[ 2\int_{0} ^{\infty} df \frac{H_g^2(f)}{f^6 H_n^2(f)} \biggr] ^{1/2} \;.
\label{eq8}
\end{equation}

Comparing Eq. (\ref{eq44}) with Eq. (\ref{eq8}) and compensating the proportional factor, and then integrating with respect to frequency, we finally obtain
\begin{eqnarray}
{\rm{SNR}}^M &=& \frac{9 H_0^2}{40 \pi^2} \sqrt{T_{\rm{obs}}} \biggl[ 2\int_{0}^{\infty} df \frac{(\Omega_{\rm{gw}}^M(f))^2 \det \mathbf{F}(f)}{f^6 {\cal{F}}_M(f)} \biggr] ^{1/2} \;, \nonumber \\
&& \label{eq72} 
\end{eqnarray}
where we redefined the Fisher matrix, $\mathbf{F}$, as the matrix
\begin{equation}
\mathbf{F} (f) \equiv \left(
\begin{array}{ccc} 
F_{TT} & F_{TV} & F_{TS} \\
F_{TV} & F_{VV} & F_{VS} \\ 
F_{TS} & F_{VS} & F_{SS}
\end{array}
\right) \;. \nonumber
\end{equation}
The quantity ${\cal{F}}_M$ is the determinant of the sub-matrix, which is constructed by removing the $M$'s elements from (new) $\mathbf{F}$.

Although so far we implicitly assume that the overlap reduction functions are time-independent, the overlap reduction functions actually depend on time in the case of a space-based detector though its orbital motion. It is easy to extend to time-dependent overlap reduction function. In Eq. (\ref{eq72}), the overlap reduction functions are included through Eqs. (\ref{eq45}) - (\ref{eq46}). As long as a stochastic GWB is stationary, the summation with respect to $i$ is equivalent to the integral with respect to time, because the correlation signals at different times can be regarded as those of the detector pairs which have different location and orientation. Therefore, Eq. (\ref{eq72}) and Eqs. (\ref{eq45}) - (\ref{eq46}) should be replaced with
\begin{eqnarray}
{\rm{SNR}}^M &=& \frac{9 H_0^2}{40 \pi^2} \biggl[ 2\int_{0}^{\infty} df \frac{(\Omega_{\rm{gw}}^M(f))^2 \det \mathbf{F}(f)}{f^6 {\cal{F}}_M(f)} \biggr] ^{1/2} \nonumber \\
&& \label{eq47} 
\end{eqnarray}
and
\begin{equation}
F_{MM^{\prime}}(f) = \sum_i \int_{0}^{T_{\rm{obs}}} dt \, \frac{\gamma_i^M (t,f) \gamma_i^{M^{\prime}} (t,f)}{{\cal{N}}_i(f) } \;, \nonumber
\end{equation}
where $M$ and $M^{\prime}$ denote polarization modes, $M,M^{\prime}=T,V,S$. 

\bibliography{GWB-nonTensor-space}

\begin{thebibliography}{65}
\expandafter\ifx\csname natexlab\endcsname\relax\def\natexlab#1{#1}\fi
\expandafter\ifx\csname bibnamefont\endcsname\relax
  \def\bibnamefont#1{#1}\fi
\expandafter\ifx\csname bibfnamefont\endcsname\relax
  \def\bibfnamefont#1{#1}\fi
\expandafter\ifx\csname citenamefont\endcsname\relax
  \def\citenamefont#1{#1}\fi
\expandafter\ifx\csname url\endcsname\relax
  \def\url#1{\texttt{#1}}\fi
\expandafter\ifx\csname urlprefix\endcsname\relax\def\urlprefix{URL }\fi
\providecommand{\bibinfo}[2]{#2}
\providecommand{\eprint}[2][]{\url{#2}}

\bibitem[{\citenamefont{Grishchuk}(1974)}]{bib41}
\bibinfo{author}{\bibfnamefont{L.~P.} \bibnamefont{Grishchuk}},
  \bibinfo{journal}{{Sov. Phys. JETP}} \textbf{\bibinfo{volume}{{\bf{40}}}},
  \bibinfo{pages}{409} (\bibinfo{year}{1974}).

\bibitem[{\citenamefont{Rubakov et~al.}(1982)\citenamefont{Rubakov, Sazhin, and
  Veryaskin}}]{bib42}
\bibinfo{author}{\bibfnamefont{V.~A.} \bibnamefont{Rubakov}},
  \bibinfo{author}{\bibfnamefont{M.~V.} \bibnamefont{Sazhin}},
  \bibnamefont{and} \bibinfo{author}{\bibfnamefont{A.~V.}
  \bibnamefont{Veryaskin}}, \bibinfo{journal}{{Phys. Lett.}}
  \textbf{\bibinfo{volume}{B {\bf{115}}}}, \bibinfo{pages}{189}
  (\bibinfo{year}{1982}).

\bibitem[{\citenamefont{Abbott and Wise}(1984)}]{bib43}
\bibinfo{author}{\bibfnamefont{L.~F.} \bibnamefont{Abbott}} \bibnamefont{and}
  \bibinfo{author}{\bibfnamefont{M.~B.} \bibnamefont{Wise}},
  \bibinfo{journal}{{Nucl. Phys.}} \textbf{\bibinfo{volume}{B {\bf{244}}}},
  \bibinfo{pages}{541} (\bibinfo{year}{1984}).

\bibitem[{\citenamefont{Allen}(1988)}]{bib44}
\bibinfo{author}{\bibfnamefont{B.}~\bibnamefont{Allen}},
  \bibinfo{journal}{{Phys. Rev.}} \textbf{\bibinfo{volume}{D {\bf{37}}}},
  \bibinfo{pages}{2078} (\bibinfo{year}{1988}).

\bibitem[{\citenamefont{Kosowsky et~al.}(1992)\citenamefont{Kosowsky, Turner,
  and Watkins}}]{bib45}
\bibinfo{author}{\bibfnamefont{A.}~\bibnamefont{Kosowsky}},
  \bibinfo{author}{\bibfnamefont{M.~S.} \bibnamefont{Turner}},
  \bibnamefont{and} \bibinfo{author}{\bibfnamefont{R.}~\bibnamefont{Watkins}},
  \bibinfo{journal}{{Phys. Rev.}} \textbf{\bibinfo{volume}{D {\bf{45}}}},
  \bibinfo{pages}{4514} (\bibinfo{year}{1992}).

\bibitem[{\citenamefont{Kamionkowski et~al.}(1994)\citenamefont{Kamionkowski,
  Kosowsky, and Turner}}]{bib46}
\bibinfo{author}{\bibfnamefont{M.}~\bibnamefont{Kamionkowski}},
  \bibinfo{author}{\bibfnamefont{A.}~\bibnamefont{Kosowsky}}, \bibnamefont{and}
  \bibinfo{author}{\bibfnamefont{M.~S.} \bibnamefont{Turner}},
  \bibinfo{journal}{{Phys. Rev.}} \textbf{\bibinfo{volume}{D {\bf{49}}}},
  \bibinfo{pages}{2837} (\bibinfo{year}{1994}).

\bibitem[{\citenamefont{Grojean and Servant}(2007)}]{bib47}
\bibinfo{author}{\bibfnamefont{C.}~\bibnamefont{Grojean}} \bibnamefont{and}
  \bibinfo{author}{\bibfnamefont{G.}~\bibnamefont{Servant}},
  \bibinfo{journal}{{Phys. Rev.}} \textbf{\bibinfo{volume}{D {\bf{75}}}},
  \bibinfo{pages}{043507} (\bibinfo{year}{2007}).

\bibitem[{\citenamefont{Kahniashvili et~al.}(2008)\citenamefont{Kahniashvili,
  Kosowsky, Gogoberidze, and Maravin}}]{bib48}
\bibinfo{author}{\bibfnamefont{T.}~\bibnamefont{Kahniashvili}},
  \bibinfo{author}{\bibfnamefont{A.}~\bibnamefont{Kosowsky}},
  \bibinfo{author}{\bibfnamefont{G.}~\bibnamefont{Gogoberidze}},
  \bibnamefont{and} \bibinfo{author}{\bibfnamefont{Y.}~\bibnamefont{Maravin}},
  \bibinfo{journal}{{Phys. Rev.}} \textbf{\bibinfo{volume}{D {\bf{78}}}},
  \bibinfo{pages}{043003} (\bibinfo{year}{2008}).

\bibitem[{\citenamefont{Khlebnikov and Tkachev}(1997)}]{bib49}
\bibinfo{author}{\bibfnamefont{S.}~\bibnamefont{Khlebnikov}} \bibnamefont{and}
  \bibinfo{author}{\bibfnamefont{I.}~\bibnamefont{Tkachev}},
  \bibinfo{journal}{{Phys. Rev.}} \textbf{\bibinfo{volume}{D {\bf{56}}}},
  \bibinfo{pages}{653} (\bibinfo{year}{1997}).

\bibitem[{\citenamefont{Easther and Lim}(2006)}]{bib50}
\bibinfo{author}{\bibfnamefont{R.}~\bibnamefont{Easther}} \bibnamefont{and}
  \bibinfo{author}{\bibfnamefont{E.~A.} \bibnamefont{Lim}},
  \bibinfo{journal}{{J. Cosmol. Astropart. Phys.}}
  \textbf{\bibinfo{volume}{{\bf{04}}}}, \bibinfo{pages}{010}
  (\bibinfo{year}{2006}).

\bibitem[{\citenamefont{Dufaux et~al.}(2007)\citenamefont{Dufaux, Bergman,
  Felder, Kofman, and Uzan}}]{bib51}
\bibinfo{author}{\bibfnamefont{J.~F.} \bibnamefont{Dufaux}},
  \bibinfo{author}{\bibfnamefont{A.}~\bibnamefont{Bergman}},
  \bibinfo{author}{\bibfnamefont{G.}~\bibnamefont{Felder}},
  \bibinfo{author}{\bibfnamefont{L.}~\bibnamefont{Kofman}}, \bibnamefont{and}
  \bibinfo{author}{\bibfnamefont{J.~P.} \bibnamefont{Uzan}},
  \bibinfo{journal}{{Phys. Rev.}} \textbf{\bibinfo{volume}{D {\bf{76}}}},
  \bibinfo{pages}{123517} (\bibinfo{year}{2007}).

\bibitem[{\citenamefont{Garcia-Bellido and Figueroa}(2007)}]{bib52}
\bibinfo{author}{\bibfnamefont{J.}~\bibnamefont{Garcia-Bellido}}
  \bibnamefont{and} \bibinfo{author}{\bibfnamefont{D.~G.}
  \bibnamefont{Figueroa}}, \bibinfo{journal}{{Phys. Rev. Lett.}}
  \textbf{\bibinfo{volume}{{\bf{98}}}}, \bibinfo{pages}{061302}
  (\bibinfo{year}{2007}).

\bibitem[{\citenamefont{J.~Garcia-Bellido and Sastre}(2008)}]{bib53}
\bibinfo{author}{\bibfnamefont{D.~G.~F.} \bibnamefont{J.~Garcia-Bellido}}
  \bibnamefont{and} \bibinfo{author}{\bibfnamefont{A.}~\bibnamefont{Sastre}},
  \bibinfo{journal}{{Phys. Rev.}} \textbf{\bibinfo{volume}{D {\bf{77}}}},
  \bibinfo{pages}{043517} (\bibinfo{year}{2008}).

\bibitem[{bib({\natexlab{a}})}]{bib6}
\bibinfo{note}{C. M. Will, {\it{Theory and experiment in gravitational
  physics}}, (Cambridge University Press (1993)}.

\bibitem[{\citenamefont{Eardley et~al.}(1973)\citenamefont{Eardley, Lee,
  Lightman, Wagoner, and Will}}]{bib5}
\bibinfo{author}{\bibfnamefont{D.~M.} \bibnamefont{Eardley}},
  \bibinfo{author}{\bibfnamefont{D.~L.} \bibnamefont{Lee}},
  \bibinfo{author}{\bibfnamefont{A.~P.} \bibnamefont{Lightman}},
  \bibinfo{author}{\bibfnamefont{R.~V.} \bibnamefont{Wagoner}},
  \bibnamefont{and} \bibinfo{author}{\bibfnamefont{C.~M.} \bibnamefont{Will}},
  \bibinfo{journal}{{Phys. Rev. Lett.}} \textbf{\bibinfo{volume}{30}},
  \bibinfo{pages}{884} (\bibinfo{year}{1973}).

\bibitem[{\citenamefont{Brans and Dicke}(1961)}]{bib36}
\bibinfo{author}{\bibfnamefont{C.}~\bibnamefont{Brans}} \bibnamefont{and}
  \bibinfo{author}{\bibfnamefont{R.~H.} \bibnamefont{Dicke}},
  \bibinfo{journal}{{Phys. Rev.}} \textbf{\bibinfo{volume}{{\bf{124}}}},
  \bibinfo{pages}{925} (\bibinfo{year}{1961}).

\bibitem[{bib({\natexlab{b}})}]{bib37}
\bibinfo{note}{Y. Fujii and K. Maeda, {\it{The Scalar-Tensor Theory of
  Gravitation}}, (Cambridge University Press (2002)}.

\bibitem[{bib({\natexlab{c}})}]{bib38}
\bibinfo{note}{T. P. Sotiriou and V. Faraoni, arXiv:0805.1726.}

\bibitem[{\citenamefont{Capozziello et~al.}(2009)\citenamefont{Capozziello,
  Laurentis, Nojiri, and Odintsov}}]{bib39}
\bibinfo{author}{\bibfnamefont{S.}~\bibnamefont{Capozziello}},
  \bibinfo{author}{\bibfnamefont{M.~D.} \bibnamefont{Laurentis}},
  \bibinfo{author}{\bibfnamefont{S.}~\bibnamefont{Nojiri}}, \bibnamefont{and}
  \bibinfo{author}{\bibfnamefont{S.~D.} \bibnamefont{Odintsov}},
  \bibinfo{journal}{{Gen. Relativ. Gravit.}}
  \textbf{\bibinfo{volume}{{\bf{41}}}}, \bibinfo{pages}{2313}
  (\bibinfo{year}{2009}).

\bibitem[{\citenamefont{Alves et~al.}(2009)\citenamefont{Alves, Miranda, and
  de~Araujo}}]{bib72}
\bibinfo{author}{\bibfnamefont{M.~E.~S.} \bibnamefont{Alves}},
  \bibinfo{author}{\bibfnamefont{O.~D.} \bibnamefont{Miranda}},
  \bibnamefont{and} \bibinfo{author}{\bibfnamefont{J.~C.~N.}
  \bibnamefont{de~Araujo}}, \bibinfo{journal}{Phys. Lett.}
  \textbf{\bibinfo{volume}{B {\bf{679}}}}, \bibinfo{pages}{401}
  (\bibinfo{year}{2009}).

\bibitem[{bib({\natexlab{d}})}]{bib73}
\bibinfo{note}{M. E. S. Alves, O. D. Miranda, and J. C. N. de Araujo,
  arXiv:1004.5580 (2010)}.

\bibitem[{\citenamefont{Dvali et~al.}(2000)\citenamefont{Dvali, Gabadadze, and
  Porrati}}]{bib40}
\bibinfo{author}{\bibfnamefont{G.}~\bibnamefont{Dvali}},
  \bibinfo{author}{\bibfnamefont{G.}~\bibnamefont{Gabadadze}},
  \bibnamefont{and} \bibinfo{author}{\bibfnamefont{M.}~\bibnamefont{Porrati}},
  \bibinfo{journal}{{Phys. Lett.}} \textbf{\bibinfo{volume}{B {\bf{485}}}},
  \bibinfo{pages}{208} (\bibinfo{year}{2000}).

\bibitem[{bib({\natexlab{e}})}]{bib61}
\bibinfo{note}{E. Komatsu and others, arXiv:1001.4538 (2010)}.

\bibitem[{bib({\natexlab{f}})}]{bib4}
\bibinfo{note}{The LIGO Scientific Collaboration and The Virgo Collaboration,
  Nature {\bf{460}}, 990 (2009)}.

\bibitem[{\citenamefont{Lee et~al.}(2008)\citenamefont{Lee, Jenet, and
  Price}}]{bib11}
\bibinfo{author}{\bibfnamefont{K.~J.} \bibnamefont{Lee}},
  \bibinfo{author}{\bibfnamefont{F.~A.} \bibnamefont{Jenet}}, \bibnamefont{and}
  \bibinfo{author}{\bibfnamefont{R.~H.} \bibnamefont{Price}},
  \bibinfo{journal}{{Astrophys. J.}} \textbf{\bibinfo{volume}{{\bf{685}}}},
  \bibinfo{pages}{1304} (\bibinfo{year}{2008}).

\bibitem[{\citenamefont{Seto et~al.}(2001)\citenamefont{Seto, Kawamura, and
  Nakamura}}]{bib7}
\bibinfo{author}{\bibfnamefont{N.}~\bibnamefont{Seto}},
  \bibinfo{author}{\bibfnamefont{S.}~\bibnamefont{Kawamura}}, \bibnamefont{and}
  \bibinfo{author}{\bibfnamefont{T.}~\bibnamefont{Nakamura}},
  \bibinfo{journal}{{Phys. Rev. Lett.}} \textbf{\bibinfo{volume}{87}},
  \bibinfo{pages}{221103} (\bibinfo{year}{2001}).

\bibitem[{bib({\natexlab{g}})}]{bib10}
\bibinfo{note}{S. Sato {\it{et al.}}, Journal of Physics: Conference Series
  {\bf{154}}, 012040 (2009)}.

\bibitem[{bib({\natexlab{h}})}]{bib9}
\bibinfo{note}{E. S. Phinney {\it{et al.}}, {\it{Big Bang Observer Mission
  Concept Study}} (NASA, 2003)}.

\bibitem[{\citenamefont{Cutler and Holz}(2009)}]{bib21}
\bibinfo{author}{\bibfnamefont{C.}~\bibnamefont{Cutler}} \bibnamefont{and}
  \bibinfo{author}{\bibfnamefont{D.~E.} \bibnamefont{Holz}},
  \bibinfo{journal}{{Phys. Rev.}} \textbf{\bibinfo{volume}{D {\bf{80}}}},
  \bibinfo{pages}{104009} (\bibinfo{year}{2009}).

\bibitem[{bib({\natexlab{i}})}]{bib19}
\bibinfo{note}{In general gravity theory, massive gravitons propagate with the
  speed less than $c$. However, the observation of the binary pulsars has
  tightly constrained the propagating speed of gravitons to be $v_g/c \geq
  0.998$ \cite{bib20}. So, we set $v_g=c$, which hardly affects the
  cross-correlation analysis.}

\bibitem[{\citenamefont{Allen and Ottewill}(1997)}]{bib62}
\bibinfo{author}{\bibfnamefont{B.}~\bibnamefont{Allen}} \bibnamefont{and}
  \bibinfo{author}{\bibfnamefont{A.~C.} \bibnamefont{Ottewill}},
  \bibinfo{journal}{Phys. Rev.} \textbf{\bibinfo{volume}{D {\bf{56}}}},
  \bibinfo{pages}{545} (\bibinfo{year}{1997}).

\bibitem[{\citenamefont{Cornish}(2001{\natexlab{a}})}]{bib63}
\bibinfo{author}{\bibfnamefont{N.~J.} \bibnamefont{Cornish}},
  \bibinfo{journal}{{Class. Quantum Grav.}}
  \textbf{\bibinfo{volume}{{\bf{18}}}}, \bibinfo{pages}{4277}
  (\bibinfo{year}{2001}{\natexlab{a}}).

\bibitem[{\citenamefont{Kudoh and Taruya}(2005)}]{bib64}
\bibinfo{author}{\bibfnamefont{H.}~\bibnamefont{Kudoh}} \bibnamefont{and}
  \bibinfo{author}{\bibfnamefont{A.}~\bibnamefont{Taruya}},
  \bibinfo{journal}{Phys. Rev.} \textbf{\bibinfo{volume}{D {\bf{71}}}},
  \bibinfo{pages}{024025} (\bibinfo{year}{2005}).

\bibitem[{\citenamefont{Taruya and Kudoh}(2005)}]{bib65}
\bibinfo{author}{\bibfnamefont{A.}~\bibnamefont{Taruya}} \bibnamefont{and}
  \bibinfo{author}{\bibfnamefont{H.}~\bibnamefont{Kudoh}},
  \bibinfo{journal}{Phys. Rev.} \textbf{\bibinfo{volume}{D {\bf{72}}}},
  \bibinfo{pages}{104015} (\bibinfo{year}{2005}).

\bibitem[{\citenamefont{Kudoh et~al.}(2006)\citenamefont{Kudoh, Taruya,
  Hiramatsu, and Himemoto}}]{bib66}
\bibinfo{author}{\bibfnamefont{H.}~\bibnamefont{Kudoh}},
  \bibinfo{author}{\bibfnamefont{A.}~\bibnamefont{Taruya}},
  \bibinfo{author}{\bibfnamefont{T.}~\bibnamefont{Hiramatsu}},
  \bibnamefont{and} \bibinfo{author}{\bibfnamefont{Y.}~\bibnamefont{Himemoto}},
  \bibinfo{journal}{Phys. Rev.} \textbf{\bibinfo{volume}{D {\bf{73}}}},
  \bibinfo{pages}{064006} (\bibinfo{year}{2006}).

\bibitem[{\citenamefont{Taruya}(2006)}]{bib67}
\bibinfo{author}{\bibfnamefont{A.}~\bibnamefont{Taruya}},
  \bibinfo{journal}{Phys. Rev.} \textbf{\bibinfo{volume}{D {\bf{74}}}},
  \bibinfo{pages}{104022} (\bibinfo{year}{2006}).

\bibitem[{\citenamefont{Thrane et~al.}(2009)\citenamefont{Thrane, Ballmer,
  Romano, Mitra, Talukder, Bose, and Mandic}}]{bib68}
\bibinfo{author}{\bibfnamefont{E.}~\bibnamefont{Thrane}},
  \bibinfo{author}{\bibfnamefont{S.}~\bibnamefont{Ballmer}},
  \bibinfo{author}{\bibfnamefont{J.~D.} \bibnamefont{Romano}},
  \bibinfo{author}{\bibfnamefont{S.}~\bibnamefont{Mitra}},
  \bibinfo{author}{\bibfnamefont{D.}~\bibnamefont{Talukder}},
  \bibinfo{author}{\bibfnamefont{S.}~\bibnamefont{Bose}}, \bibnamefont{and}
  \bibinfo{author}{\bibfnamefont{V.}~\bibnamefont{Mandic}},
  \bibinfo{journal}{Phys. Rev.} \textbf{\bibinfo{volume}{D {\bf{80}}}},
  \bibinfo{pages}{122002} (\bibinfo{year}{2009}).

\bibitem[{\citenamefont{Drasco and Flanagan}(2003)}]{bib69}
\bibinfo{author}{\bibfnamefont{S.}~\bibnamefont{Drasco}} \bibnamefont{and}
  \bibinfo{author}{\bibfnamefont{E.~E.} \bibnamefont{Flanagan}},
  \bibinfo{journal}{Phys. Rev.} \textbf{\bibinfo{volume}{D {\bf{67}}}},
  \bibinfo{pages}{082003} (\bibinfo{year}{2003}).

\bibitem[{\citenamefont{Himemoto et~al.}(2007)\citenamefont{Himemoto, Taruya,
  Kudoh, and Hiramatsu}}]{bib70}
\bibinfo{author}{\bibfnamefont{Y.}~\bibnamefont{Himemoto}},
  \bibinfo{author}{\bibfnamefont{A.}~\bibnamefont{Taruya}},
  \bibinfo{author}{\bibfnamefont{H.}~\bibnamefont{Kudoh}}, \bibnamefont{and}
  \bibinfo{author}{\bibfnamefont{T.}~\bibnamefont{Hiramatsu}},
  \bibinfo{journal}{Phys. Rev.} \textbf{\bibinfo{volume}{D {\bf{75}}}},
  \bibinfo{pages}{022003} (\bibinfo{year}{2007}).

\bibitem[{\citenamefont{Seto}(2008)}]{bib71}
\bibinfo{author}{\bibfnamefont{N.}~\bibnamefont{Seto}},
  \bibinfo{journal}{Astrophys. J.} \textbf{\bibinfo{volume}{683}},
  \bibinfo{pages}{L95} (\bibinfo{year}{2008}).

\bibitem[{\citenamefont{Maggiore}(2000)}]{bib3}
\bibinfo{author}{\bibfnamefont{M.}~\bibnamefont{Maggiore}},
  \bibinfo{journal}{{Phys. Rep.}} \textbf{\bibinfo{volume}{{\bf{331}}}},
  \bibinfo{pages}{283} (\bibinfo{year}{2000}).

\bibitem[{\citenamefont{Allen and Romano}(1999)}]{bib27}
\bibinfo{author}{\bibfnamefont{B.}~\bibnamefont{Allen}} \bibnamefont{and}
  \bibinfo{author}{\bibfnamefont{J.~D.} \bibnamefont{Romano}},
  \bibinfo{journal}{{Phys. Rev.}} \textbf{\bibinfo{volume}{D {\bf{59}}}},
  \bibinfo{pages}{102001} (\bibinfo{year}{1999}).

\bibitem[{bib({\natexlab{j}})}]{bib1}
\bibinfo{note}{If a theory contains a parity-violating term, $+$ and $\times$
  modes are polarized. Such a model predicts a GWB with circular polarizations.
  The detectability has been discussed in \cite{bib14,bib18,bib28,bib29}.}

\bibitem[{\citenamefont{Christensen}(1992)}]{bib25}
\bibinfo{author}{\bibfnamefont{N.}~\bibnamefont{Christensen}},
  \bibinfo{journal}{{Phys. Rev.}} \textbf{\bibinfo{volume}{D {\bf{46}}}},
  \bibinfo{pages}{5250} (\bibinfo{year}{1992}).

\bibitem[{\citenamefont{Flanagan}(1993)}]{bib26}
\bibinfo{author}{\bibfnamefont{E.~E.} \bibnamefont{Flanagan}},
  \bibinfo{journal}{{Phys. Rev.}} \textbf{\bibinfo{volume}{D {\bf{48}}}},
  \bibinfo{pages}{2389} (\bibinfo{year}{1993}).

\bibitem[{\citenamefont{Nishizawa et~al.}(2009)\citenamefont{Nishizawa, Taruya,
  Hayama, Kawamura, and Sakagami}}]{bib12}
\bibinfo{author}{\bibfnamefont{A.}~\bibnamefont{Nishizawa}},
  \bibinfo{author}{\bibfnamefont{A.}~\bibnamefont{Taruya}},
  \bibinfo{author}{\bibfnamefont{K.}~\bibnamefont{Hayama}},
  \bibinfo{author}{\bibfnamefont{S.}~\bibnamefont{Kawamura}}, \bibnamefont{and}
  \bibinfo{author}{\bibfnamefont{M.}~\bibnamefont{Sakagami}},
  \bibinfo{journal}{{Phys. Rev.}} \textbf{\bibinfo{volume}{{D \bf{79}}}},
  \bibinfo{pages}{082002} (\bibinfo{year}{2009}).

\bibitem[{bib({\natexlab{k}})}]{bib16}
\bibinfo{note}{That is calculated with DECIGO design parameters, assuming the
  noise curve is quantum-noise limited. The parameters we used are the arm
  length $1000\,\rm{km}$, the angular frequency of a laser $3.6\times
  10^{15}\,\rm{sec}^{-1}$, the laser power $10\,\rm{W}$, the mirror mass
  $100\,{\rm{kg}}$, and the finesse of the cavity 10.}

\bibitem[{\citenamefont{Seto and Taruya}(2008)}]{bib14}
\bibinfo{author}{\bibfnamefont{N.}~\bibnamefont{Seto}} \bibnamefont{and}
  \bibinfo{author}{\bibfnamefont{A.}~\bibnamefont{Taruya}},
  \bibinfo{journal}{{Phys. Rev.}} \textbf{\bibinfo{volume}{D {\bf{77}}}},
  \bibinfo{pages}{103001} (\bibinfo{year}{2008}).

\bibitem[{\citenamefont{Holz and Hughes}(2005)}]{bib24}
\bibinfo{author}{\bibfnamefont{D.~E.} \bibnamefont{Holz}} \bibnamefont{and}
  \bibinfo{author}{\bibfnamefont{S.~A.} \bibnamefont{Hughes}},
  \bibinfo{journal}{Astrophys. J.} \textbf{\bibinfo{volume}{{\bf{629}}}},
  \bibinfo{pages}{15} (\bibinfo{year}{2005}).

\bibitem[{\citenamefont{Seto}(2007)}]{bib18}
\bibinfo{author}{\bibfnamefont{N.}~\bibnamefont{Seto}},
  \bibinfo{journal}{{Phys. Rev.}} \textbf{\bibinfo{volume}{D {\bf{75}}}},
  \bibinfo{pages}{061302(R)} (\bibinfo{year}{2007}).

\bibitem[{\citenamefont{Farmer and Phinney}(2003)}]{bib8}
\bibinfo{author}{\bibfnamefont{A.~J.} \bibnamefont{Farmer}} \bibnamefont{and}
  \bibinfo{author}{\bibfnamefont{E.~S.} \bibnamefont{Phinney}},
  \bibinfo{journal}{{Mon. Not. R. Astron. Soc.}}
  \textbf{\bibinfo{volume}{{\bf{346}}}}, \bibinfo{pages}{1197}
  (\bibinfo{year}{2003}).

\bibitem[{\citenamefont{Friedman et~al.}(2006)\citenamefont{Friedman, Cooray,
  and Melchiorri}}]{bib54}
\bibinfo{author}{\bibfnamefont{B.~C.} \bibnamefont{Friedman}},
  \bibinfo{author}{\bibfnamefont{A.}~\bibnamefont{Cooray}}, \bibnamefont{and}
  \bibinfo{author}{\bibfnamefont{A.}~\bibnamefont{Melchiorri}},
  \bibinfo{journal}{{Phys. Rev.}} \textbf{\bibinfo{volume}{D {\bf{74}}}},
  \bibinfo{pages}{123509} (\bibinfo{year}{2006}).

\bibitem[{\citenamefont{Chongchitnan and Efstathiou}(2006)}]{bib55}
\bibinfo{author}{\bibfnamefont{S.}~\bibnamefont{Chongchitnan}}
  \bibnamefont{and}
  \bibinfo{author}{\bibfnamefont{G.}~\bibnamefont{Efstathiou}},
  \bibinfo{journal}{{Phys. Rev.}} \textbf{\bibinfo{volume}{D {\bf{73}}}},
  \bibinfo{pages}{083511} (\bibinfo{year}{2006}).

\bibitem[{\citenamefont{Smith et~al.}(2006)\citenamefont{Smith, Peiris, and
  Cooray}}]{bib56}
\bibinfo{author}{\bibfnamefont{T.~L.} \bibnamefont{Smith}},
  \bibinfo{author}{\bibfnamefont{H.~V.} \bibnamefont{Peiris}},
  \bibnamefont{and} \bibinfo{author}{\bibfnamefont{A.}~\bibnamefont{Cooray}},
  \bibinfo{journal}{{Phys. Rev.}} \textbf{\bibinfo{volume}{D {\bf{73}}}},
  \bibinfo{pages}{123503} (\bibinfo{year}{2006}).

\bibitem[{\citenamefont{Saito and Yokoyama}(2009)}]{bib58}
\bibinfo{author}{\bibfnamefont{R.}~\bibnamefont{Saito}} \bibnamefont{and}
  \bibinfo{author}{\bibfnamefont{J.}~\bibnamefont{Yokoyama}},
  \bibinfo{journal}{{Phys. Rev. Lett.}} \textbf{\bibinfo{volume}{{\bf{102}}}},
  \bibinfo{pages}{161101} (\bibinfo{year}{2009}).

\bibitem[{\citenamefont{Armstrong et~al.}(1999)\citenamefont{Armstrong,
  Estabrook, and Tinto}}]{bib33}
\bibinfo{author}{\bibfnamefont{J.~W.} \bibnamefont{Armstrong}},
  \bibinfo{author}{\bibfnamefont{F.~B.} \bibnamefont{Estabrook}},
  \bibnamefont{and} \bibinfo{author}{\bibfnamefont{M.}~\bibnamefont{Tinto}},
  \bibinfo{journal}{Astrophys. J.} \textbf{\bibinfo{volume}{527}},
  \bibinfo{pages}{814} (\bibinfo{year}{1999}).

\bibitem[{\citenamefont{Cornish}(2001{\natexlab{b}})}]{bib34}
\bibinfo{author}{\bibfnamefont{N.~J.} \bibnamefont{Cornish}},
  \bibinfo{journal}{{Phys. Rev.}} \textbf{\bibinfo{volume}{D {\bf{65}}}},
  \bibinfo{pages}{022004} (\bibinfo{year}{2001}{\natexlab{b}}).

\bibitem[{\citenamefont{Prince et~al.}(2002)\citenamefont{Prince, Tinto,
  Larson, and Armstrong}}]{bib35}
\bibinfo{author}{\bibfnamefont{T.~A.} \bibnamefont{Prince}},
  \bibinfo{author}{\bibfnamefont{M.}~\bibnamefont{Tinto}},
  \bibinfo{author}{\bibfnamefont{S.~L.} \bibnamefont{Larson}},
  \bibnamefont{and} \bibinfo{author}{\bibfnamefont{J.~W.}
  \bibnamefont{Armstrong}}, \bibinfo{journal}{{Phys. Rev.}}
  \textbf{\bibinfo{volume}{D {\bf{66}}}}, \bibinfo{pages}{122002}
  (\bibinfo{year}{2002}).

\bibitem[{\citenamefont{Schilling}(1997)}]{bib30}
\bibinfo{author}{\bibfnamefont{R.}~\bibnamefont{Schilling}},
  \bibinfo{journal}{{Class. Quantum Grav.}}
  \textbf{\bibinfo{volume}{{\bf{14}}}}, \bibinfo{pages}{1513}
  (\bibinfo{year}{1997}).

\bibitem[{\citenamefont{Rakhmanov}(2008)}]{bib31}
\bibinfo{author}{\bibfnamefont{M.}~\bibnamefont{Rakhmanov}},
  \bibinfo{journal}{{Class. Quantum Grav.}}
  \textbf{\bibinfo{volume}{{\bf{25}}}}, \bibinfo{pages}{184017}
  (\bibinfo{year}{2008}).

\bibitem[{\citenamefont{Nishizawa et~al.}(2008)}]{bib32}
\bibinfo{author}{\bibfnamefont{A.}~\bibnamefont{Nishizawa}}
  \bibnamefont{et~al.}, \bibinfo{journal}{{Phys. Rev.}}
  \textbf{\bibinfo{volume}{D {\bf{77}}}}, \bibinfo{pages}{022002}
  (\bibinfo{year}{2008}).

\bibitem[{bib({\natexlab{l}})}]{bib20}
\bibinfo{note}{L. S. Finn and P. J. Sutton, Phys. Rev. D {\bf{65}}, 044022
  (2002).}

\bibitem[{\citenamefont{Seto}(2006)}]{bib28}
\bibinfo{author}{\bibfnamefont{N.}~\bibnamefont{Seto}},
  \bibinfo{journal}{{Phys. Rev. Lett.}} \textbf{\bibinfo{volume}{{\bf{97}}}},
  \bibinfo{pages}{151101} (\bibinfo{year}{2006}).

\bibitem[{\citenamefont{Seto and Taruya}(2007)}]{bib29}
\bibinfo{author}{\bibfnamefont{N.}~\bibnamefont{Seto}} \bibnamefont{and}
  \bibinfo{author}{\bibfnamefont{A.}~\bibnamefont{Taruya}},
  \bibinfo{journal}{{Phys. Rev. Lett.}} \textbf{\bibinfo{volume}{{\bf{99}}}},
  \bibinfo{pages}{121101} (\bibinfo{year}{2007}).

\bibitem[{\citenamefont{Will}(2006)}]{bib13}
\bibinfo{author}{\bibfnamefont{C.~M.} \bibnamefont{Will}},
  \bibinfo{journal}{{Living Rev. Relativity}} \textbf{\bibinfo{volume}{9}},
  \bibinfo{pages}{3} (\bibinfo{year}{2006}).

\end{thebibliography}

\end{document}